\documentclass[twocolumn]{revtex4-1}

\usepackage{latexsym}
\usepackage{amssymb}
\usepackage{amsmath}
\usepackage{amsthm}
\usepackage{amsfonts}
\usepackage{ifthen}
\usepackage{revsymb}
\usepackage{yfonts}
\usepackage{graphicx}
\usepackage{algpseudocode}
\usepackage{bbm}
\usepackage{multirow}
\usepackage{gensymb}
\usepackage{epstopdf}
\usepackage{color}
\usepackage{afterpage}
\usepackage{verbatim}
\usepackage{soul}
\usepackage{lineno, blindtext}
\usepackage{footnote}
\usepackage{afterpage}
\usepackage{algpseudocode}
\usepackage[normalem]{ulem}
\usepackage{algorithm}
\usepackage{algpseudocode}
\usepackage[colorlinks = true]{hyperref}
\usepackage{longtable}

\usepackage{xcolor}
\definecolor{darkred}  {rgb}{0.5,0,0}
\definecolor{darkblue} {rgb}{0,0,0.5}
\definecolor{darkgreen}{rgb}{0,0.5,0}

\hypersetup{
  urlcolor   = blue,         
  linkcolor  = darkblue,     
  citecolor  = darkgreen,    
  filecolor  = darkred       
}

\newcommand{\be}{\begin{equation}}
\newcommand{\ee}{\end{equation}}
\newcommand{\bq}{\begin{eqnarray}}
\newcommand{\eq}{\end{eqnarray}}
\newcommand{\bea}{\begin{eqnarray}}
\newcommand{\eea}{\end{eqnarray}}
\newcommand{\ba}{\begin{align}}
\newcommand{\ea}{\end{align}}

\newcommand{\ket}[1]{ | \, #1 \rangle}
\newcommand{\bra}[1]{ \langle #1 \,  |}

\newcommand{\bZ}{\mathbbm{Z}}

\newcommand{\cO}{\mathcal O}

\definecolor{mygray}{gray}{0.6}

\newcommand{\beginsupplement}{%
        \setcounter{table}{0}
        \renewcommand{\thetable}{S\arabic{table}}%
        \setcounter{figure}{0}
        \renewcommand{\thefigure}{S\arabic{figure}}%
     }

\definecolor{mygray}{gray}{0.9}

\newcommand{\emp}[1]{{\color{blue} #1}}

\begin{document}

\title{ Hardware-efficient Variational Quantum Eigensolver for Small Molecules and Quantum Magnets}

\author{Abhinav Kandala}
\thanks{These authors contributed equally to this work.}
\author{Antonio Mezzacapo}
\thanks{These authors contributed equally to this work.}
\author{Kristan Temme}
\author{\mbox{Maika Takita} }
\author{Markus Brink }
\author{Jerry M. Chow}
\author{Jay M. Gambetta }\affiliation{IBM T.J.  Watson  Research  Center,  Yorktown  Heights,  NY 10598,  USA}

\date{\today}

\maketitle
\noindent

\textbf{Quantum computers can be used to address molecular structure, materials science and condensed matter physics problems, which currently stretch the limits of existing high-performance computing resources~\cite{NERSCReport2015}. Finding exact numerical solutions to these interacting fermion problems has exponential cost, while Monte Carlo methods  are plagued by the fermionic sign problem. These limitations of classical computational methods have made even few-atom molecular structures problems of practical interest for medium-sized quantum computers. Yet, thus far experimental implementations have been restricted to molecules involving only Period I elements~ \cite{Lanyon2010,Du10,Peruzzo13,Wang15,OMalley16,Shen17,Paesani2017}. Here, we demonstrate the experimental optimization of up to six-qubit Hamiltonian problems with over a hundred Pauli terms, determining the ground state energy for molecules of increasing size, up to $\textrm{BeH}_2$. This is enabled by a hardware-efficient variational quantum eigensolver with trial states specifically tailored to the available interactions in our quantum processor, combined with a compact encoding of fermionic Hamiltonians ~\cite{Tapering} and a robust stochastic optimization routine~\cite{Spall1992}. We further demonstrate the flexibility of our approach by applying the technique to a problem of quantum magnetism~\cite{Lanyon2011}. Across all studied problems, we find agreement between experiment and numerical simulations with a noisy model of the device. These results help elucidate the requirements for scaling the method to larger systems, and aim at bridging the gap between problems at the forefront of high-performance computing and their implementation on quantum hardware.}

The fundamental goal of addressing molecular structure problems is to solve for the ground state energy of many-body interacting fermionic Hamiltonians. Solving this problem on a quantum computer relies on a mapping between fermionic and qubit operators~\cite{BK2002}. This restates it as a specific instance of a local Hamiltonian problem on a set of qubits. Given a $k$-local Hamiltonian $H$, composed of terms that act on at most $k$ qubits, the solution to the local Hamiltonian problem amounts to finding its smallest eigenvalue $E_G$, 
\begin{equation}
\label{HamProblem}
H\ket{\Phi}=E_G\ket{\Phi}.
\end{equation}
To date, no efficient algorithm is known that can solve this problem in full generality. For $k\geq2$ the problem is known to be QMA-complete~\cite{Kempe06}. However, it is expected that physical systems have Hamiltonians that do not constitute hard instances of this problem, and can be solved efficiently on a quantum computer, while remaining hard to solve classically. 

Following Feynman's idea for quantum simulation, a quantum algorithm for the ground state problem of interacting fermions was proposed in~\cite{Abrams97} and~\cite{Guzik05}. The approach relies on a good initial state that has a large overlap with the ground state and then solves the problem using the quantum phase estimation algorithm (PEA)~\cite{KitaevPhase}. While PEA has been demonstrated to achieve extremely accurate energy estimates for quantum chemistry~\cite{Lanyon2010,Du10,Wang15,Paesani2017}, it applies stringent requirements on quantum coherence. 

\begin{figure*}
\includegraphics[width=6in]{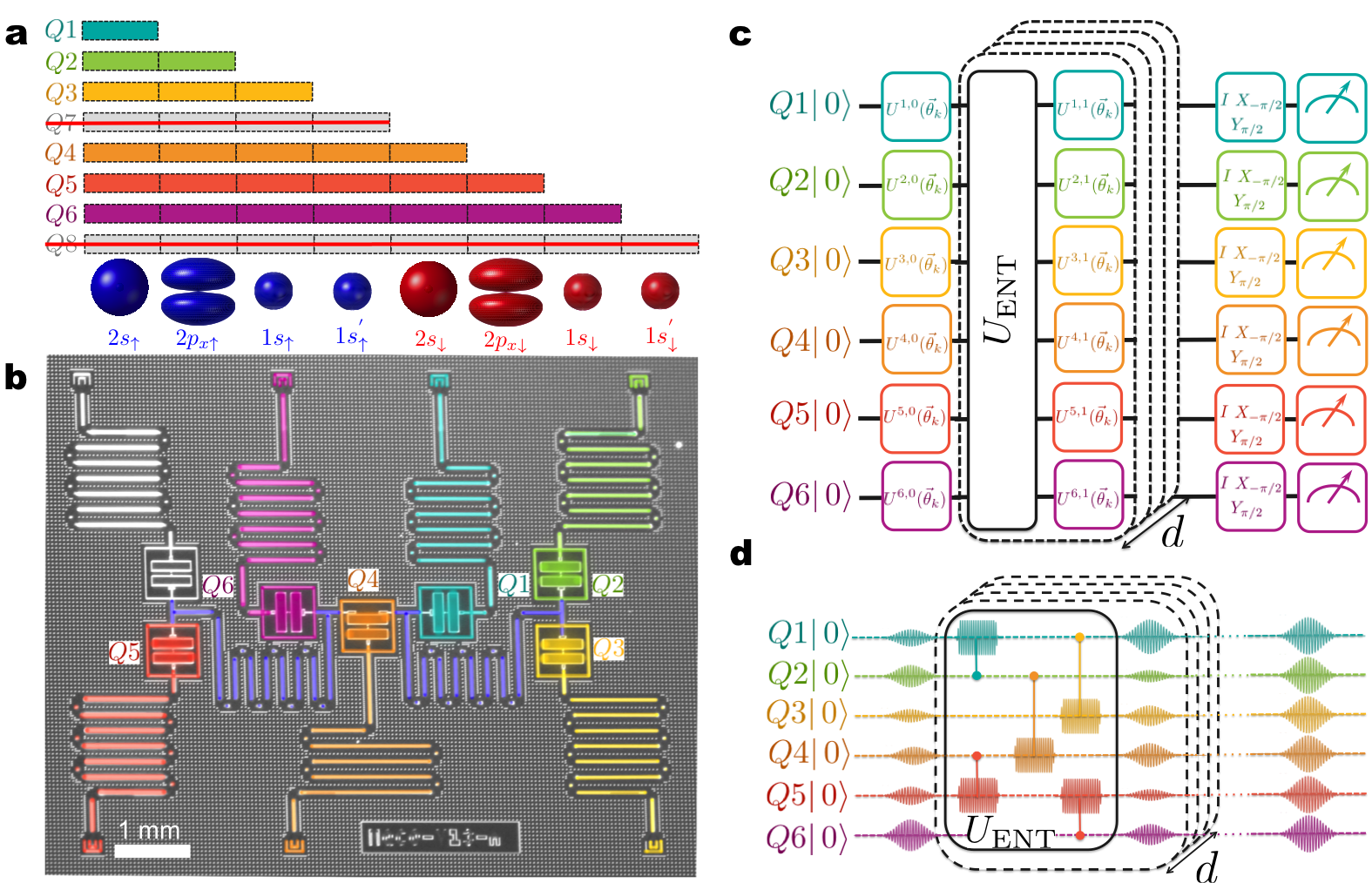}
\caption{\label{Figure1}   \textbf{Quantum chemistry on a superconducting quantum processor: device and quantum circuit for variational trial state preparation.}  Solving molecular structure problems on a quantum computer relies on mappings between fermionic and qubit operators. \textbf{a} Parity mapping of 8 spin orbitals (drawn in blue and red, not to scale) onto 8 qubits, reduced to 6 qubits via qubit tapering of fermionic spin-parity symmetries. The bars indicate the parity of the spin-orbitals encoded in each qubit. {\bf b} False colored optical micrograph of the superconducting quantum processor. The transmon qubits are coupled via two CPW resonators, highlighted in blue, and have individual CPW resonators for control and readout. \textbf{c} Hardware-efficient quantum circuit for trial state preparation and energy estimation, shown here for 6 qubits. The circuit is composed of a sequence of interleaved single-qubit rotations, and entangling unitary operations $U_\textrm{\textrm{ENT}}$ that entangle all the qubits in the circuit. A final set of post-rotations prior to qubit readout are used to measure the expectation values of the terms in the qubit Hamiltonian, and estimate the energy of the trial state. \textbf{d} An example of the pulse sequence for the preparation of a six qubit trial state, where $U_\textrm{\textrm{ENT}}$ is implemented as a sequence of two-qubit cross resonance gates.} 
\end{figure*}

An alternative approach is the use of quantum optimizers. Their utility spans from combinatorial optimization problems~\cite{farhi2014quantum,Farhi2017} to quantum chemistry in the form of variational quantum eigensolvers (VQEs), where they were introduced to reduce coherence requirements on quantum hardware~\cite{Peruzzo13,yung2013transistor,mcclean2016theory}. The VQE uses Ritz's variational principle to prepare approximations to the ground state and its energy. In this approach, the quantum computer is used to prepare variational trial states that depend on a set of parameters. Then, the expectation value of the energy is estimated and used by a classical optimizer to generate a new set of improved parameters. The advantage of VQE over classical simulation
methods is that is can prepare trial states that are not amenable to efficient classical numerics.

To date, the VQE approach realized in experiment has been limited by different factors. Typically, one considers a unitary coupled cluster (UCC) ansatz for the trial state \emp{\cite{OMalley16,Shen17}}, which has a number of parameters that scale quartically with the number of spin-orbitals considered, in the single and double excitation approximation. Furthermore, when implementing the UCC ansatz on a quantum computer, one has to account for Trotterization errors~\cite{mcclean2016theory,Wecker2015,Romero2017}. In this work, we introduce and implement a ``hardware-efficient" ansatz preparation for VQE, where trial states are parameterized by quantum gates that are tailored to the physical device available. We numerically show the viability of such trial states for small molecular structure problems and use a superconducting quantum processor to perform optimizations of the molecular energies of $\textrm{H}_2$, $\textrm{LiH}$ and $\textrm{BeH}_2$, and extend its application to a Heisenberg antiferromagnetic model in an external magnetic field.

The device used in the experiments is a superconducting quantum processor with six fixed-frequency transmon qubits, together with a central weakly-tunable asymmetric transmon qubit~\cite{hutchings2017tunable}. The device is cooled down in a dilution refrigerator, thermally anchored to its mixing chamber plate at 25 mK. The experiments discussed here make use of six of these qubits (labeled Q1-6), highlighted in Fig.~\ref{Figure1}b. The qubits are coupled via two superconducting coplanar waveguide (CPW) resonators that serve as quantum buses, and can be individually controlled and read out through independent readout resonators.

The hardware-efficient trial states we consider use the naturally available entangling interactions of the superconducting hardware, described by a drift Hamiltonian $H_0$ that generates the entanglers $U_{\textrm{ENT}} = \exp(-i H_0 \tau)$ which entangle all the qubits in the circuit. These are interleaved with arbitrary single-qubit Euler rotations which are implemented  as a combination of  $Z$ and $X$ gates, given by $U^{q,i}(\vec\theta) = Z^q_{\theta^{q,i}_1}X^q_{\theta^{q,i}_2}Z^q_{\theta^{q,i}_3}$, where $q$ identifies the qubit and $i=0,1,...d$ refers to the depth position, as depicted in Fig.~\ref{Figure1}c. The $N$-qubit trial states are obtained from the state $\ket{00\ldots 0}$, applying $d$ entanglers $U_{\textrm{ENT}}$ that alternate with $N$ Euler rotations, giving

\begin{align}
\ket{\Phi(\vec{\theta})}=&\prod_{q=1}^N\left[ U^{q,d}(\vec{\theta})\right]\times U_{\textrm{\textrm{ENT}}}\times \prod_{q=1}^N\left[ U^{q,d-1}(\vec{\theta})\right]\nonumber\\
\label{VarState} & \cdots\times  U_{\textrm{\textrm{ENT}}}\times  \prod_{q=1}^N\left[ U^{q,0}(\vec{\theta})\right]\ket{00...0}.
\end{align}

Since the qubits are all initialized in their ground state $\ket{0}$, the first set of Z rotations of $U^{q,0}(\vec{\theta})$ is not implemented, resulting in a total of $p=N(3d+2)$ independent angles. In the experiment, the evolution time $\tau$ and the individual couplings in $H_0$ can be controlled. However, numerical simulations indicate that accurate optimizations are obtained for fixed-phase $U_{\textrm{\textrm{ENT}}}$, leaving the $p$ control angles as variational parameters . Our hardware-efficient approach does not rely on the accurate implementation of specific two qubit gates and can be used with any $U_\textrm{\textrm{ENT}}$ that generates sufficient entanglement. This is in contrast to UCC trial states that require high-fidelity quantum gates approximating a unitary operator tailored on a theoretical ansatz. For the experiments considered here, the entanglers $U_\textrm{\textrm{ENT}}$ are composed of a sequence of two-qubit cross-resonance (CR) gates~\cite{Sheldon2016}.Simulations as a function of entangler phase show plateaus of minimal energy error around gate phases corresponding to the maximal pairwise concurrence, see Supplementary Information. We therefore set  the entangler evolution time $\tau$ at the beginning of such plateaus, in order to reduce decoherence effects.

\begin{figure}
\includegraphics[width=3.5in]{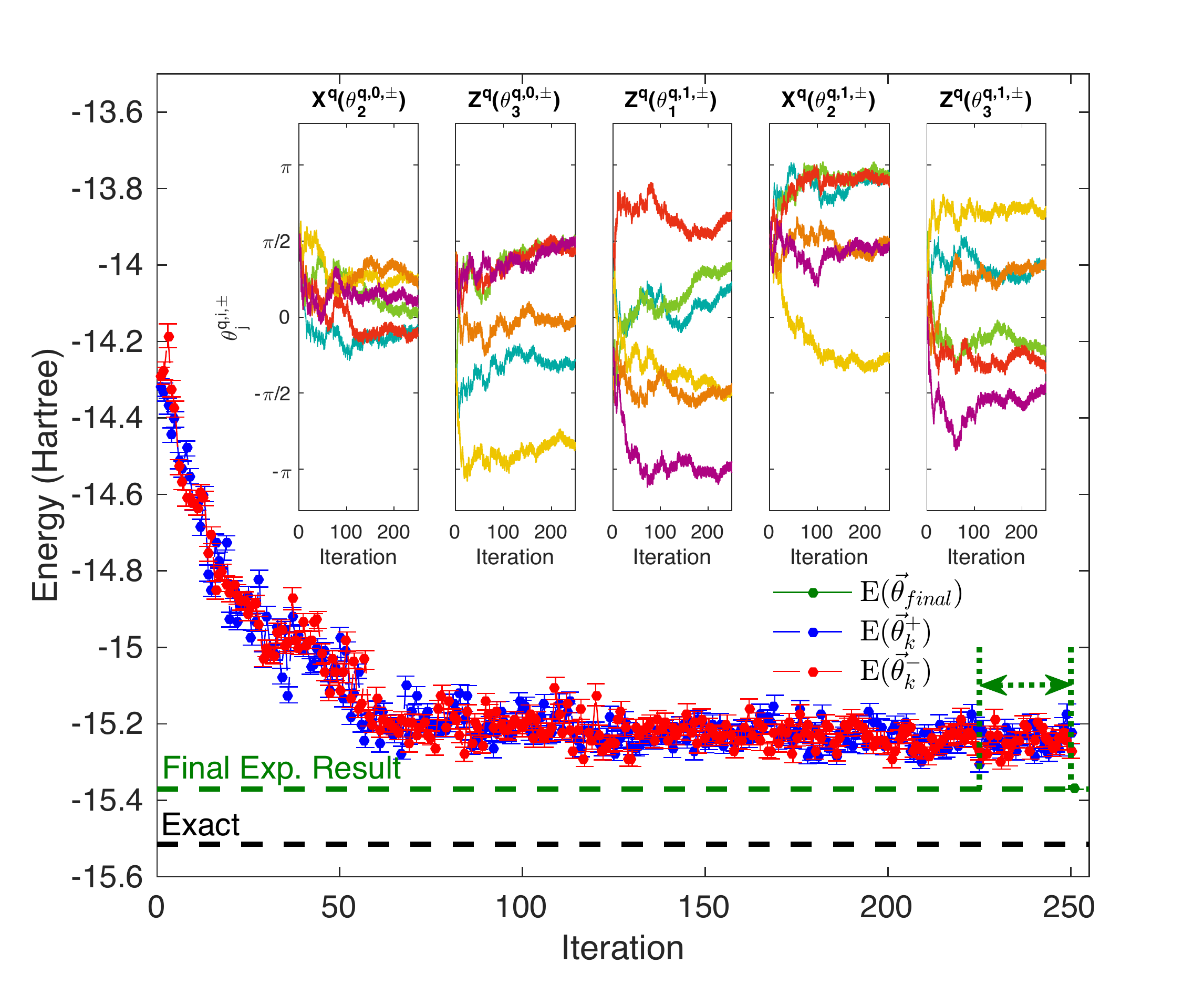}
\caption{\label{Figure2} \textbf{Experimental implementation of six-qubit optimization.} Energy minimization for the six-qubit Hamiltonian describing  BeH$_{2}$ at interatomic distance $l=1.7 \buildrel _{\circ} \over {\mathrm{A}}$, plotted against the exact value (black dashed line). For each iteration $k$, the gradient at each control $\vec{\theta_k}$ is approximated using $10^3$ samples for energy estimations at $\vec{\theta_k^+}$ and $\vec{\theta_k^-}$, shown in blue and red, respectively. The inset shows the simultaneous optimization of 30 Euler angles that control the trial state preparation. Each color refers to a particular qubit, following the qubit color scheme of Fig.~\ref{Figure1}. The final energy estimate (green dashed line) is obtained using the angles $\vec{\theta_{\textrm{final}}}$, averaged over the last $25$ angle updates, in order to mitigate the effect of stochastic fluctuations, with a higher number of $10^5$ samples, to get a more accurate energy estimation.}
\end{figure}

In our experiments, the $Z$ rotations are implemented as frame changes in the control software~\cite{mckay2016efficient}, while the $X$ rotations are implemented by appropriately scaling the amplitude of calibrated $X_\pi$ pulses, using a fixed total time of $100$~ns for every single-qubit rotation. The CR$_{c-t}$ gates that compose $U_{\mathrm{\textrm{\textrm{ENT}}}}$ are implemented by driving a control qubit Q$_c$ with a microwave pulse resonant with a target qubit Q$_t$. Hamiltonian tomography of the CR$_{c-t}$ gates is used to reveal the strengths of the various interaction terms, and the gate time for maximal entanglement~\cite{Sheldon2016}. We set our two-qubit gate times at $150$~ns, simultaneously trying to minimize the effect of decoherence without compromising the accuracy of the optimization outcome, see Supplementary Information.

After each trial state is prepared, we estimate the associated energy by measuring the expectation values of the individual Pauli terms in the Hamiltonian. These estimates are affected by stochastic fluctuations due to finite sampling. Different post-rotations are applied after trial state preparation for sampling different Pauli operators, see Fig.~\ref{Figure1}c,d. We group the Pauli operators into tensor product basis sets that require the same post-rotations. We numerically show that such grouping reduces the energy fluctuations, keeping the same total number of samples, thereby reducing the time overhead for energy estimation, see Supplementary Information. The energy estimates are then used by a gradient descent algorithm that relies on a simultaneous perturbation stochastic approximation (SPSA) to update the control parameters. The SPSA algorithm approximates the gradient using only two energy measurements, regardless of the dimensions of the parameter space $p$, achieving a level of accuracy comparable to standard gradient descent methods, in the presence of stochastic fluctuations~\cite{Spall1992}. This is a crucial aspect for optimizing over many qubits and long depths for trial state preparation, allowing us to optimize over a number of parameters as large as $p=30$.

\begin{figure*}
\includegraphics[width=7in]{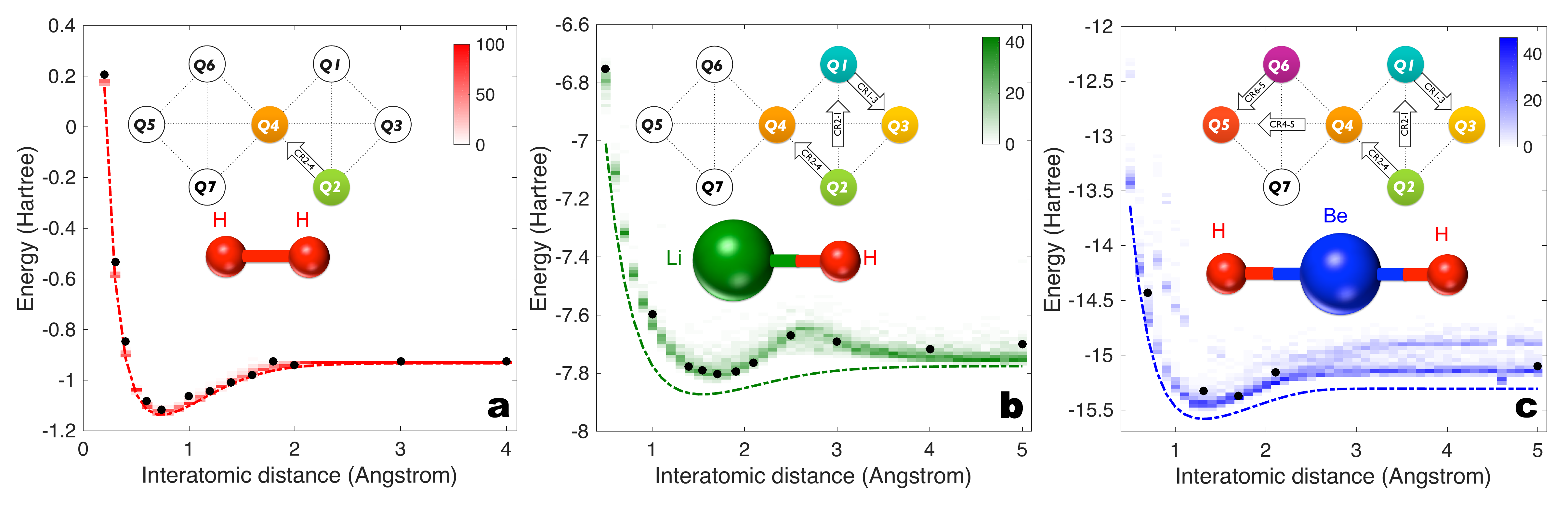}
\caption{\label{Figure3}  \textbf{Application to quantum chemistry: Potential energy surfaces } Experimental results (black circles), exact energy surfaces (dotted lines) and density plots of outcomes from numerical simulations, for a number of interatomic distances for \textbf{a}, H$_{2}$ \textbf{b}, LiH, and \textbf{c}, BeH$_{2}$. The experimental and numerical results presented here use depth $d=1$ circuits. The error bars on the experimental data are smaller than the size of the markers. The density plots are obtained from 100 numerical outcomes at each interatomic distance. The top insets of each figure highlight the qubits used for the experiment, and the cross-resonance gates that constitute $U_\textrm{\textrm{ENT}}$. The bottom insets of each figure are representations of the molecular geometry, not drawn to scale. For all the three molecules, the deviation of the experimental results from the exact curves, is well explained by the stochastic simulations.}
\end{figure*}

To address molecular problems on our quantum processor, we rely on a compact encoding of the second-quantized fermionic Hamiltonians on to qubits. The $\textrm{H}_2$ molecular Hamiltonian has $4$ spin-orbitals, representing the spin-degenerate $1s$ orbitals of the two Hydrogen atoms. We use a binary tree encoding~\cite{BK2002} to map it to a $4$ qubit system, and remove two qubits associated with the spin-parities of the system~\cite{Tapering}. The $\textrm{BeH}_2$ Hamiltonian is defined upon the $1s$, $2s$, $2p_x$ orbitals associated to Be, and $1s$ orbital associated to each H atom, for a total of $10$ spin orbitals. We then assume perfect filling of the two innermost $1s$ spin-orbitals of Be, after dressing them via the diagonalization of the non-interacting part of the fermionic Hamiltonian. We map the 8 spin-orbital Hamiltonian of BeH$_2$ spin-orbital Hamiltonian using the parity mapping, and remove, as in the case of $\textrm{H}_2$, two qubits associated to the spin-parity symmetries, reducing this to a $6$ qubit problem that encodes $8$ spin-orbitals. A similar approach is also used to map LiH onto $4$ qubits. The Hamiltonians for $\textrm{H}_2$, $\textrm{LiH}$ and $\textrm{BeH}_2$ at their equilibrium distance are explicitly given in the Supplementary Information.

The results from an optimization procedure are illustrated in detail in Fig.~\ref{Figure2}, using the BeH$_{2}$ Hamiltonian for the interatomic distance of $1.7 \buildrel _{\circ} \over {\mathrm{A}}$. It is important to note that while using a large number of entanglers $U_\textrm{ENT}$ helps achieve better energy estimates in the absence of noise, the combined effect of decoherence and finite sampling sets the optimal depth for optimizations on our quantum hardware to $0-2$ entanglers. The results presented in Fig.~\ref{Figure2} are obtained using a depth $d=1$ circuit, with a total of $30$ Euler control angles associated with $6$ qubits. The inset of Fig.~\ref{Figure2} shows the simultaneous perturbation of 30 Euler angles, as the energy estimates are updated. 

To obtain the potential energy surfaces for H$_{2}$, LiH, and BeH$_{2}$, we search for the ground state energy of their molecular Hamiltonians, using 2, 4, and 6 qubits respectively, for depth $d=1$, for a range of different interatomic distances. The experimental results are compared with the ground state energies obtained from exact diagonalization and outcomes from numerical simulations in Fig.~\ref{Figure3}. The colored density plots in each panel are obtained from $100$ numerical optimizations for each interatomic distance, using CR entangling gates on the same topology as the experiments. These numerics account for decoherence effects, simulated by adding amplitude damping and dephasing channels after each layer of quantum gates. The impact of finite sampling on the optimization algorithm is taken into account by numerically sampling the individual Pauli terms in the Hamiltonian, and adding their averages. The strengths of the noise channels are derived from the measured values for $T_1$, $T_2^{*}$ coherence times. In addition to the effects of decoherence and noisy energy estimates, the deviations are also due to low circuit depth for trial state preparation, which, for example, explains the kink in the range $l=2.5-3 \buildrel _{\circ} \over {\mathrm{A}}$, in Fig.~\ref{Figure3}b. In the absence of noise, critical depths of $d=1,8,28 (1,6,16)$ are required to achieve chemical accuracy (approx. 0.0016 Hartree), on the current experimental (all-to-all) connectivities for H$_2$, LiH and BeH$_2$, respectively, see Supplementary Information. In contrast, a generic UCC ansatz truncated to the second order for a 8-orbital molecule such as our model of BeH$_2$ would require 4160 fermionic variational terms, which, after accounting for fermionic mappings and Trotterization would generate a number of quantum gates of the same order. The scaling of resources and noise requirements to achieve chemical accuracy using hardware-efficient trial states are detailed in the Supplementary Information. We emphasize that our approach is unaffected by coherent gate errors, which shifts the focus to the reduction of incoherent errors, favoring our fixed-frequency, all-microwave control, qubit architecture. Furthermore, the effect of incoherent errors can be mitigated as recently proposed~\cite{Mitigation,Mitigation2,Mitigation3}, without requiring additional quantum resources.

\begin{figure}
\includegraphics[width=3in]{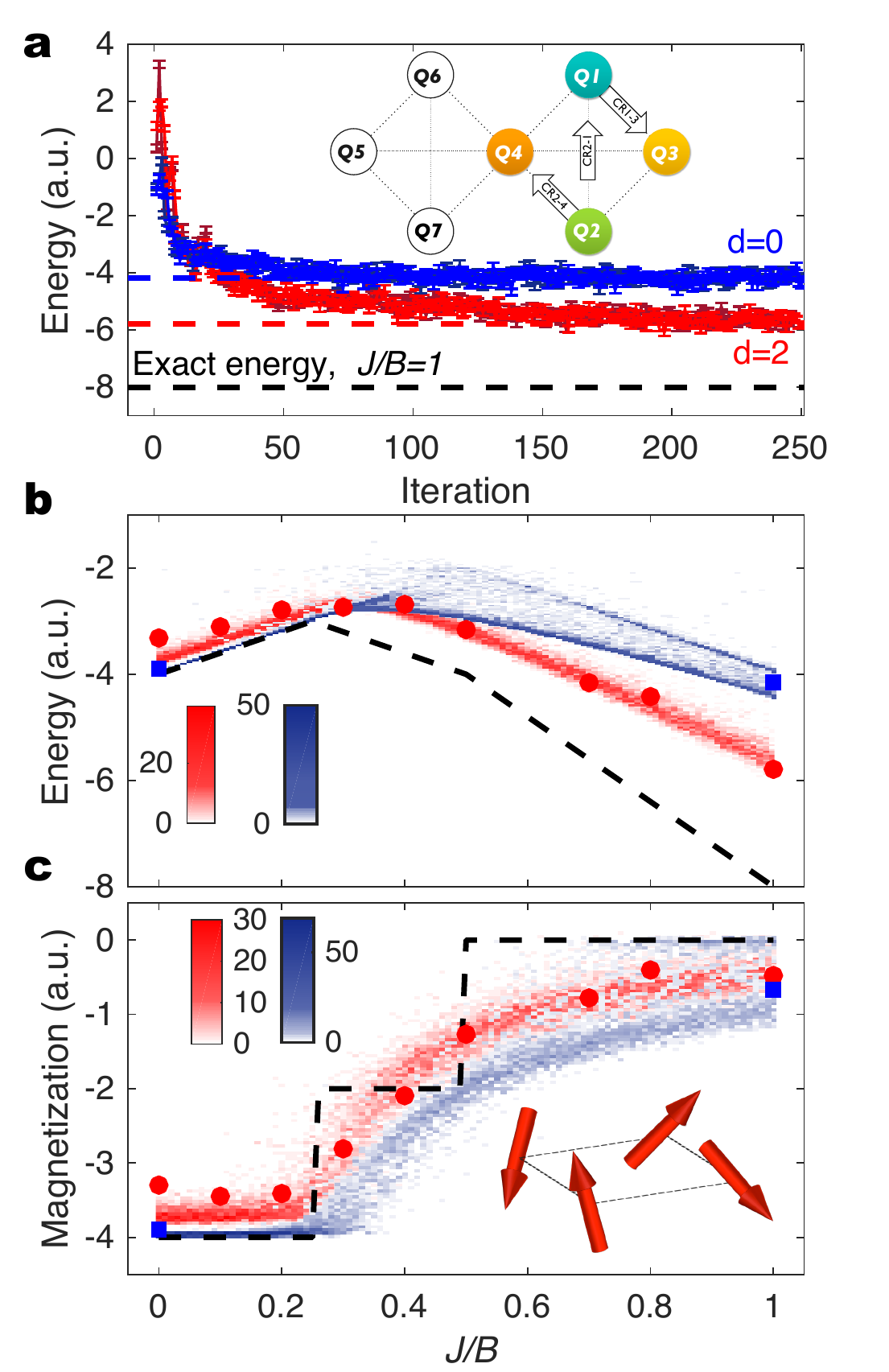}
\caption{\label{Figure4} \textbf{Application to quantum magnetism: 4 qubit Heisenberg model on a square lattice, in an external magnetic field.} Comparison of the optimization using $d=0$ (blue) and $d=2$ (red) circuits for state preparation. \textbf{a} Energy optimization for $J/B=1$, plotted against the exact energy (dashed black line). The inset of highlights the qubits used for the experiment, and the cross-resonance gates that constitute $U_\textrm{\textrm{ENT}}$. Experimental results for $d=0$ (blue squares) and $d=2$ (red circles) plotted against exact curves (black dashed lines) and density plots of 100 numerical outcomes, for {\bf b} energy and {\bf c} magnetization, for a range of $J/B$ ratios.}
\end{figure}

We now demonstrate the applicability of our technique to a problem of quantum magnetism, and show that with the same noisy quantum hardware, the advantage of using higher circuits depths is crucially dependent on the target Hamiltonian. Specifically, we consider a four qubit Heisenberg model on a square lattice, in the presence of an external magnetic field. The model is described by the Hamiltonian $H=J\sum_{\langle{ij}\rangle}(X_iX_j+Y_iY_j+Z_iZ_j)+B\sum_{i}Z_i$, where $\langle{ij}\rangle$ indicates the nearest neighbor pairs, $J$ is the strength of the spin-spin interaction, and $B$ the magnetic field along the $Z$-direction. We utilize our technique to solve for the ground state energy of the system for a range of $J/B$ values. When $J=0$, the ground state is completely separable, and the best estimates are obtained for depth $d=$0. As $J$ is increased, the ground state is increasingly entangled, and the best estimates are instead obtained at $d=2$, despite the increased decoherence caused by using two entanglers for trial state preparation. This is shown in Fig.~\ref{Figure4}a for $J/B=1$. The experimental results are compared with the exact ground state energies for a range of $J/B$ values in Fig.~\ref{Figure4}b, and our deviations are captured by the density plots of the numerical outcomes that account for noisy energy estimations and decoherence. Furthermore, in Fig.~\ref{Figure4}c, we show that our approach can also be used to evaluate observables such as the magnetization of the system $M_z$.

The experiments presented here have shown that a hardware-efficient VQE implemented on a six-qubit superconducting quantum processor is capable of addressing molecular problems beyond period 1 elements, up to BeH$_2$. A numerical analysis for the hardware requirements to improve the accuracy of a VQE for the molecules addressed suggest the need for dramatic improvements in coherence and sampling, see Supplementary Information. For more complex problems, increased coherence and faster gates would enable longer circuit depths for state preparation while an increased on-chip qubit connectivity is crucial for reducing critical depth requirements.The use of fast reset schemes~\cite{Bultnick2016} would enable increased sampling rates, improving the effectiveness of the classical optimizer, while reducing time overheads. The performance of the quantum-classical feedback loop could be further improved by variants~\cite{AdaptiveSPSA} of the simultaneous perturbation protocol discussed here. Trial state preparation circuits, combining better ansatzes from classical approximate methods and hardware-efficient gates, can be further investigated to improve on the current ansatzes. Finally, in the absence of a fault tolerant architechture, the agreement of our experimental results with the noise models considered opens a path to error mitigation protocols for experimentally accessible circuit depths~\cite{Mitigation,Mitigation2,Mitigation3}.

{\bf Supplementary Information} is available in the online version of the paper.

{\bf Acknowledgments}
We thank J. Chavez-Garcia, A. D. Corcoles and J. Rozen for experimental contributions; J. Hertzberg and S. Rosenblatt for room temperature characterization; B. Abdo for providing the Jospehson Parametric Converters;  S. Brayvi, J. Smolin, E. Magesan, L. Bishop, S. Sheldon, N. Moll, P. Barkoutsos, and I. Tavernelli for valuable discussions; W. Shanks for assistance with the experimental control software. We thank A. D. Corcoles for edits to the manuscript. We acknowledge support from the IBM Research Frontiers Institute. We  acknowledge  support  from  IARPA  under  contract  W911NF-10-1-0324 for device fabrication.

{\bf Author contributions}
A.K. and A.M. contributed equally to this work. J.M.G and K.T designed the experiments. A.K and M.T characterized the device and A.K performed the the experiments. M.B fabricated the devices. AM developed the theory and the numerical simulations. A.K, A.M and J.M.G interpreted and analyzed the experimental data. A.K, A.M, K.T, J.M.C and J.M.G contributed to the composition of the manuscript. 

{\bf Author information} The authors declare no competing financial interests. Correspondence and requests for materials should be addressed to A.K. (akandala@us.ibm.com) or A.M. (amezzac@us.ibm.com)

\pagebreak

\onecolumngrid

\section*{ Supplementary Information: Hardware-efficient Quantum Optimizer for Small Molecules and Quantum Magnets}
\beginsupplement

\section{Device and characterization}

\begin{figure*}[b]
\includegraphics[width=4.5in]{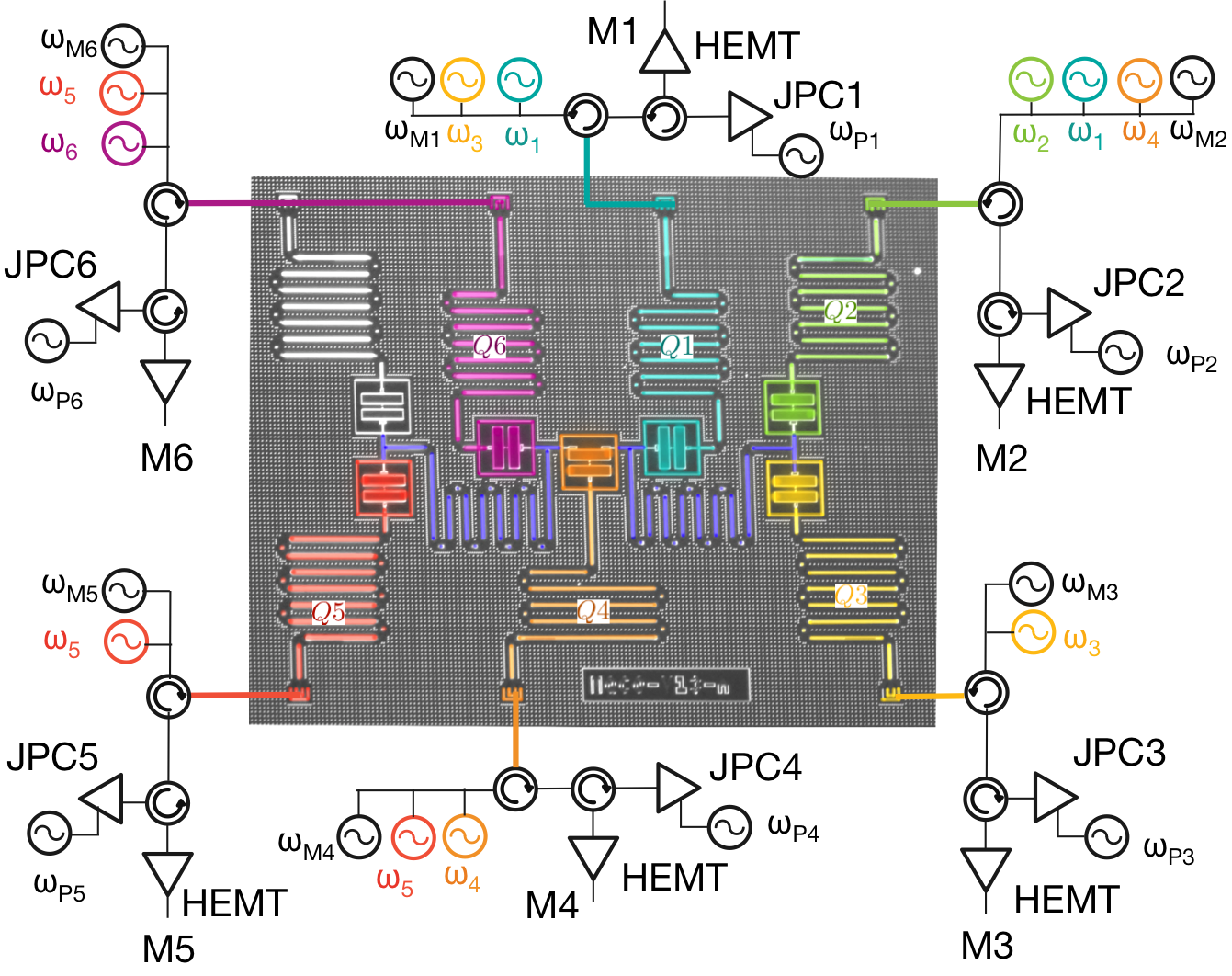}
\caption{\label{Circuit} {\bf Device and circuit schematic} False colored optical micrograph depicts the components of our superconducting quantum processor: seven transmon qubits, two shared CPW resonators (in blue) for qubit-qubit coupling, and seven individual CPW resonators used for both, qubit control and readout. The qubits are controlled solely by microwave pulses that are delivered from the room temperature electronics via attenuated coaxial lines. The single qubit gates are implemented by microwave drives at the specific qubit $Q_i$'s frequency $\omega_i$ , while the entangling two-qubit CR gates are implemented by driving a control qubit $Q_c $ at the frequency $\omega_t$ of the target qubit $Q_t$, where $i,c,t \in \{1,2,3,4,5,6\}$. The state of each qubit is measured at its readout resonator frequency $\omega_{Mi}$. The reflected readout signals are amplified first by a JPC, pumped at a frequency $\omega_{Pi}$, followed by HEMT amplifiers at 4K. }
\end{figure*}

The fundamental building blocks of our quantum hardware are superconducting Josephson junction (JJ) based qubits. The physical device includes 6 fixed frequency transmon qubits and a central flux-tunable asymmetric transmon qubit ~\cite{hutchings2017tunableS}. For the experiments discussed in this paper, we use 6 of these qubits, including the central flux-tunable qubit. The device connectivity is provided by two superconducting coplanar waveguide (CPW) resonators acting as quantum information buses, each of which couples four qubits, with the central asymmetric transmon coupled to both buses (see Fig.~\ref{Circuit}). Each qubit has its own individual CPW resonator for control and readout. The device is fabricated on a Si wafer using a single step of photolithography and sputtering for the superconducting Nb resonators and qubit capacitor pads, followed by e-beam lithography and double angle evaporation to define the Al-based JJ's. Refer to ~\cite{Chow2014S,Corcoles2015S} for further fabrication details.

\begin{figure*}
\includegraphics[width=3in]{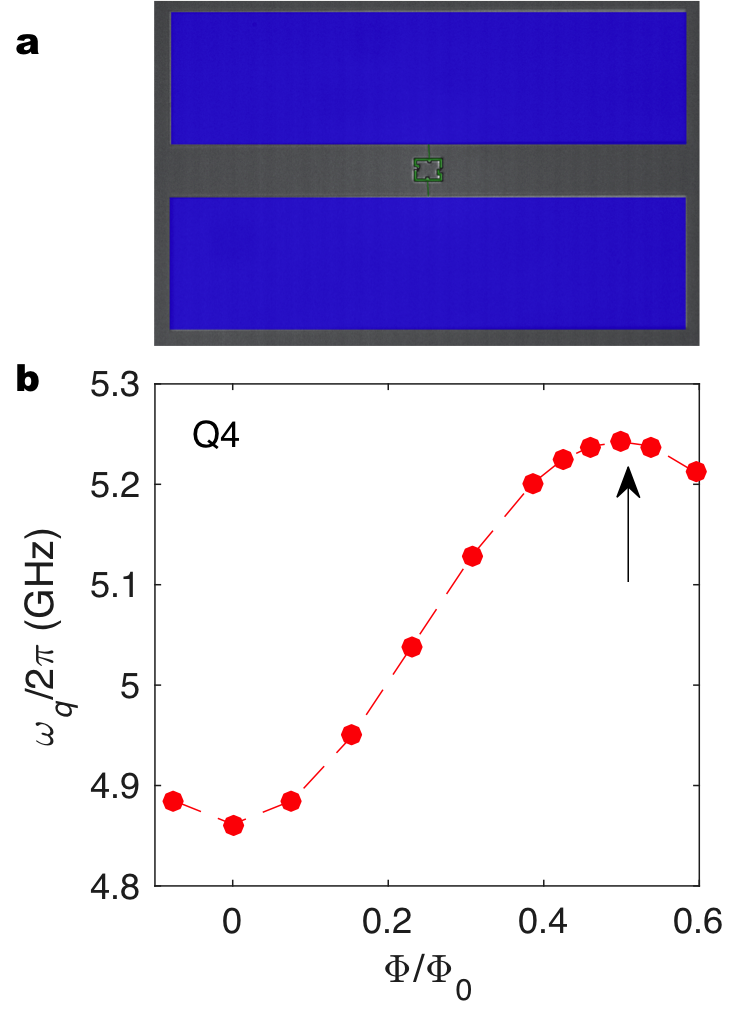}
\caption{\label{Supp_Asymmon} {\bf Asymmetric transmon and tuning curve} {\bf a} False-colored optical micrograph of an asymmetric transmon, with an Al SQUID loop (in green), shunted by Nb capacitor pads (in blue). {\bf b} Qubit frequency versus flux for the asymmetric transmon Q4. A constant flux offset is subtracted, and the flux is expressed in units of the flux quantum $\Phi_0=h/2e$, where $h$ is Planck's constant, and $e$ is electric charge. The qubit is operated at its upper flux sweet spot, indicated by the arrow. The dashed line is a guide to the eye.}
\end{figure*}

Frequency crowding is an important issue for large networks of fixed frequency qubits employing cross resonance (CR) as an entangling gate, leading to crosstalk, leakage out of the computational sub-space or very slow gate times. Furthermore, current fabrication capabilities make it challenging to control the frequencies of transmons to within 200 MHz. In this context, we designed our central qubit Q4, which is directly coupled to all other qubits on the chip, to be weakly frequency tunable for reduced sensitivity to flux noise~\cite{hutchings2017tunableS}. The qubit is referred to as an `asymmetric transmon', and uses a superconducting quantum interference device (SQUID) as its inductive element. The two junctions in the SQUID however have different Jospehson energies, engineered by varying the size of the junctions. An external superconducting coil is used to tune Q4 to its upper sweet spot, which, in the current experiment, is the optimal point for CR gates to its neighbors. The flux-tuning curve is shown in Fig.~\ref{Supp_Asymmon}. At its upper sweet spot, Q4 is operated as a fixed frequency transmon, with coherence times that are comparable to other qubits on the chip Table~\ref{table:device_parameter}.

The qubits are readout by dispersive measurements through independent readout resonators, with each readout line having a sequence of low temperature amplifiers --- a Josephson parametric converter (JPC) ~\cite{Bergeal2010S,Abdo2011S} followed by a high electron mobility transistor (model : LNF-LNC4\textunderscore8A) --- for achieving high assignment fidelity. For a measurement time of 1.5 $\mu$s, the joint readout assignment errors on Q2, Q4, Q6 are $<0.06$, and $<0.03$ for Q1, Q3, and Q5. The anharmonicity of the fixed frequency qubits are $\sim$ 310 MHz, while the asymmetric transmon has an anharmonicity of $\sim$ 300 MHz. Further details of the device parameters are listed in Table~\ref{table:device_parameter}. 

\begin{table}
	\begin{tabular*}{6.5in}{l|@{\extracolsep{\fill}}*{6}{c}}
		\hline \hline
		Qubit 		           & $\text{Q}_1$ 	    & $\text{Q}_2$ 	    & $\text{Q}_3$ 	    & $\text{Q}_4$ 	    & $\text{Q}_5$ 	    & $\text{Q}_6$	\\ \hline
		$\omega_{01}/2\pi$  (GHz)&5.3206		&	5.3567		&	5.2926		&5.2455		& 5.2999         &5.3882
		\\ \hline
		$T_1$($\mu$s)  & $24.7\pm3.2 $  & $42.0\pm5.1$  & $20.4\pm4.4$  & $42.3\pm5.2$   & $44.4\pm4.9$   &$20.6\pm0.8$
		\\ \hline
		$T_{2}$($\mu$s)  & $31.1\pm6.1 $  & $38.7\pm12.5$  & $35.3\pm8.7$  & $47.4\pm14.0$  &   $60.5\pm8.7$  &$26.4\pm4.3$
		\\ \hline
		$T_2^*$($\mu$s)  & $22.2\pm4.8 $  & $28.6\pm1.2$  & $6.2\pm0.9$  & $36.7\pm10.5$  &   $40.0\pm3.2$     &$27\pm2.8$
		\\ \hline
		$\omega_{r}/2\pi$  (GHz) &	6.6223		& 6.6892		&	6.5589	& 6.7154		& 6.6532     &6.5885 
		\\ \hline
		$\delta/2\pi$ (GHz) &-0.311	&-0.312	&-0.315	&-0.299	&-0.311	&-0.310
		\\ \hline
		$\epsilon_{r}$  &	0.0240		& 0.0544		&	0.0291	& 	0.0469	&  0.0278    &0.0507
		\\ \hline \hline 					
	\end{tabular*}
		\caption{\label{table:device_parameter} \textbf{Qubit and readout characterization.} Qubit transitions ($\omega_{01}/2\pi$), average relaxation times ($T_1$), average coherence times ($T_2$, $T_2^*$), readout resonator frequencies ($\omega_{r}/2\pi$), qubit anharmonicity ($\delta/2\pi$), readout assignment errors ($\epsilon_{r}$) for the six qubits discussed in the paper.}	
\end{table}

\begin{figure*}
\includegraphics[width=6in]{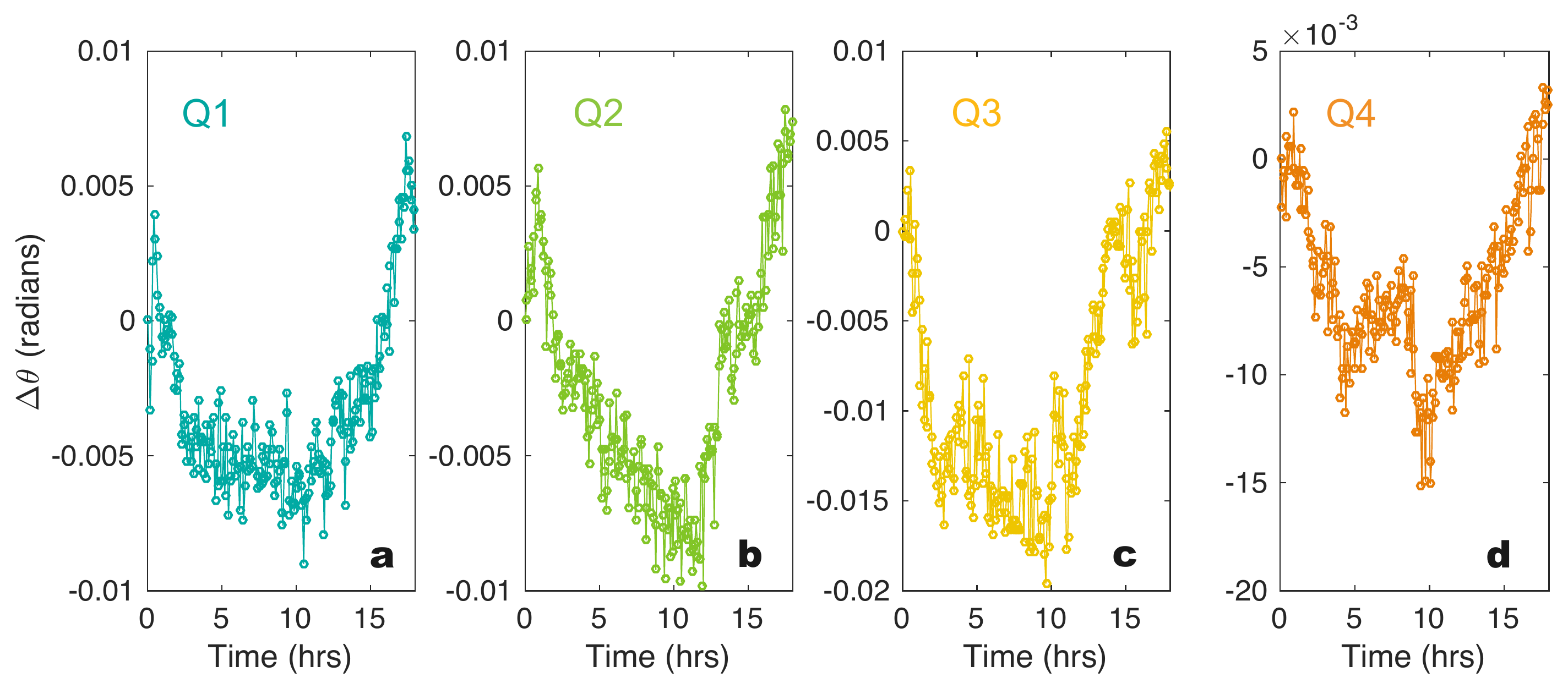}
\caption{\label{Supp_sqgdrift} {\bf Single qubit gate drifts} Repeated calibrations of the amplitude for a $X_{\pi}$ pulse over 18 hours for Q1-4 ({\bf a-d}) reveal the magnitude and timescale for drifts in the amplitude of the single qubit gates. Here, the amplitude drifts are scaled as angle deviations $\Delta\theta$ from the starting $X_{\pi}$-rotation.}
\end{figure*}

The experimental implementation of variational quantum algorithms requires stability of the gates used for trial state preparation. Given the long times associated with optimization of large Hamiltonians, we periodically calibrate the amplitude and phase of our single-qubit and two-qubit gates during the course of the experiment. In order to estimate the time scale and magnitude of drifts in pulse amplitude and phase, we repeatedly calibrate our gates over several hours. For instance, Fig.~\ref{Supp_sqgdrift} shows the drifts in the pulse amplitude for calibrated $X_{\pi}$ pulses, expressed as angle deviations from the starting 180$^o$ $X$-rotation. Over the course of 18 hours, the deviations are less than $1.5^o$.

\section{Hardware-efficient optimization of quantum Hamiltonian problems\label{Scheme}}

We present here a compact scheme describing the whole optimization algorithm. The individual subroutines of the method will be described in the following sections. 
\begin{algorithm}[H]
\caption{Hardware-efficient optimization of quantum Hamiltonian problems}
\begin{algorithmic}[1]
\State Map the quantum Hamiltonian problem to a qubit Hamiltonian $H$ 
\State Choose a depth $d$ for the quantum circuit that prepares the trial state
\State Choose a set of variational controls $\vec\theta_1$ that parametrize the starting trial state 
\State Choose a number of samples $S$ for the feedback loop and one $S_f$ for the final estimation
\State Choose a number of maximal control updates $k_L$
\While{$E_f$ has not converged}
\Procedure{Quantum Feedback Loop}{}
\For{\texttt{$k=1$ to $k_L$ }}
	 	\State Prepare trial states around $\vec\theta_k$ and evaluate $\langle H \rangle$ with $S$ samples 
	 	\State Update and store the controls $\vec\theta_k$
\EndFor
\State Evaluate $E_f=\langle H \rangle$ using the best controls with $S_f$ samples
\EndProcedure
\State Increase $d$, $k_L$, $S$, $S_f$
\EndWhile
\State{\Return $E_f$}
\end{algorithmic}
\end{algorithm}
In the above algorithm, the first item describes the encoding of quantum Hamiltonians on a set of qubits. In the case of addressing a fermionic problem, we use an encoding and qubit reduction scheme from Ref.~\cite{TaperingS}, explained in Section~\ref{HamDerivation}, which is convenient for the molecular problems considered in this work. In general, different encodings could be considered, such as ones based on first-quantization methods. The outcome of the optimization depends on the parameters $d, k_L, S, S_f$, and in general will be better as these are increased, up to a point in which either one saturates the quantum resources available (e.g. decoherence limit, sampling time), or the optimization outcome $E_f$ has converged: in this case increasing $d, k_L, S, S_f$ will not improve the final answer $E_f$. In Section~\ref{EntSect} we describe the specific entangling gates we have used int the experiment to prepare trial states. In Section~\ref{AppEnergyEst} we give details on the evaluation of the mean energy $\langle H\rangle$, and its dependence on the total number of samples $S$ and experimental assignement errors. The energies measured in this way are then fed to a classical optimizer, described in Section~\ref{SPSA}. In Section~\ref{Scaling} we numerically estimate the resources required (circuit depth $d$, number of control updates $k_L$, number of samples $S$) to improve the accuracy of the optimization outcome.

\section{Molecular Hamiltonians\label{HamDerivation}}

The molecular Hamiltonians considered in this work are computed in the STO-3G basis, using the software PyQuante~\cite{PyQuanteS} to obtain the one and two-electron integrals. The STO-3G minimal basis is obtained by fitting three gaussians to the Slater atomic orbitals, and commonly used in quantum chemistry because of the efficiency in obtaining electronic integrals~\cite{szabo1989modernS}. For the H$_2$ molecule, each atom contributes a 1s orbital, for a total of 4 spin-orbitals. We set the X axis as the interatomic axis for the LiH and BeH$_2$ molecules, and consider the orbitals 1s for each H atom and 1s, 2s, 2p$_x$ for the Li and Be atoms, assuming zero filling for the 2p$_y$ and 2p$_z$ orbitals, which do not interact strongly with the subset of orbitals considered. This choice of orbitals amounts to a total of 8 spin-orbitals for LiH and 10 for BeH$_2$. The Hamiltonians are expressed using the second quantization language,
\begin{align}
\label{Htgt}
H =H_1+H_2=\sum_{\alpha, \beta=1}^M t_{\alpha \beta} \, a^\dag_{\alpha} a_{\beta} +\frac{1}{2}  \sum_{\alpha, \beta, \gamma, \delta = 1}^M u_{\alpha \beta \gamma \delta}\, a^\dag_{\alpha} a^\dag_{\gamma} a_{\delta} a_{\beta},
\end{align}
where $a^\dag_\alpha$($a_\alpha$) is the fermionic creation(annihilation) operator of the fermionic mode $\alpha$, satisfying fermionic commutation rules $\{a_\alpha,a_\beta\}=0$, $\{a^\dag_\alpha,a^\dag_\beta\}=0$, $\{a_\alpha, a^\dag_\beta\}=\delta_{\alpha\beta}$. Here $M=4,8,10$ is the number of spin-orbitals for H$_2$, LiH and BeH$_2$ respectively, and we have used the chemists' notation~\cite{szabo1989modernS} for the two-body integrals,
\begin{align}
t_{\alpha\beta}&=\int d\vec x_1\Psi_\alpha(\vec{x}_1) \left(-\frac{\vec\nabla_1^2}{2}+\sum_{i} \frac{Z_i}{|\vec{r}_{1i}|}\right)\Psi_\beta (\vec{x}_1),\\
u_{\alpha\beta\gamma\delta}&=\int\int d \vec{x}_1 d \vec{x}_2 \Psi_\alpha^*(\vec{x}_1)\Psi_\beta(\vec{x}_1)\frac{1}{|\vec{r}_{12}|}\Psi_\gamma^*(\vec{x}_2)\Psi_\delta(\vec{x}_2),
\end{align}
where we have defined the nuclei charges $Z_i$, the nuclei-electron and electron-electron separations $\vec{r}_{1i}$ and $\vec{r}_{12}$, the $\alpha$-th orbital wavefunction $\Psi_\alpha(\vec{x}_1)$, and we have assumed that the spin is conserved in the spin-orbital indices $\alpha,\beta$ and $\alpha,\beta,\gamma,\delta$.
In the case of LiH and BeH$_2$, we then consider perfect filling for the inner 1s orbitals, dressed in the basis in which $H_1$ is diagonal. To this extent, we first implement a Bogoliubov transformation on the modes $a'_\alpha=\sum_\beta U_{\alpha\beta}a_\beta $, such that 
\begin{equation}
H_1^d=U^\dag H_1U,\hspace{0.5cm}H^d_1=\sum_{\alpha=1}^M \omega'_\alpha a'^\dag_\alpha a'_\alpha.
\end{equation}
We then consider the ``dressed'' 1s modes of Li and Be to be filled, efficiently obtaining an effective Hamiltonian acting on generic states of the form $\ket{\Psi}=a'^\dag_{1s\uparrow}a'^\dag_{1s\downarrow}\left(\sum_{\beta\neq1s\sigma}\psi_\beta a'^\dag_{\beta}\right)\ket{0}$, where $\psi_\beta$ are generic normalized coefficients, and $1s\sigma=\{1s\uparrow,1s\downarrow\}$ refers to the inner 1s orbitals of Li and Be. Note that this approximation is valid whenever $-\omega'_{1s\sigma}\gg |u'_{\alpha\beta\gamma\delta}|\hspace{0.2cm}\forall\sigma,\alpha,\beta,\gamma,\delta$, i.e. in the case of very low-energy orbitals that do not interact strongly with the higher-energy ones. The ansatz $\ket{\Psi}=a'^\dag_{1s\uparrow}a'^\dag_{1s\downarrow}\left(\sum_{\beta\neq1s\sigma}\psi_\beta a'^\dag_{\beta}\right)\ket{0}$ allows to define an effective screened Hamiltonian on the 1s orbitals for the hydrogen atoms, and 2s and 2p$_x$ for Lithium and Berillium, for a total of 6 and 8 spin-orbitals for LiH and BeH$_2$, respectively. According to this ansatz, the one-body fermionic terms containing the filled orbitals will now contribute as a shift to the total energy (I here is the identity operator)
\be
\omega'_{1\uparrow}a'^\dag_{1\uparrow} a'_{1\uparrow} \rightarrow\omega'_{1\uparrow} \textrm{I},\hspace{0.5cm}\omega'_{1\downarrow}a'^\dag_{1\downarrow} a'_{1\downarrow} \rightarrow\omega'_{1\downarrow} \textrm{I},
\ee
while some of the two-body interactions, containing the set $F$ of 1s filled modes of Li and Be, $F=\{1s\uparrow,1s\downarrow\}$, become effective one-body or energy shift terms,
\begin{align}
\frac{u'_{\alpha\beta\gamma\delta}}{2} a'^\dag_{\alpha} a'^\dag_{\gamma} a'_{\delta}a'_\beta\rightarrow
\begin{cases}
\frac{u'_{\alpha\beta\gamma\delta}}{2}a'^\dag_\gamma a'_\delta,\hspace{0.5cm}&\alpha=\beta, \alpha\in F,\{\gamma,\delta\}\notin F\\
\frac{u'_{\alpha\beta\gamma\delta}}{2}a'^\dag_\alpha a'_\beta,\hspace{0.5cm}&\gamma=\delta, \gamma\in F,\{\alpha,\beta\}\notin F\\
-\frac{u'_{\alpha\beta\gamma\delta}}{2}a'^\dag_\gamma a'_\beta,\hspace{0.5cm}&\alpha=\delta, \alpha\in F,\{\beta,\gamma\}\notin F\\
-\frac{u'_{\alpha\beta\gamma\delta}}{2}a'^\dag_\alpha a'_\delta,\hspace{0.5cm}&\gamma=\beta, \gamma\in F,\{\alpha,\delta\}\notin F\\
\frac{u'_{\alpha\beta\gamma\delta}}{2}I,\hspace{0.5cm}&\alpha=\beta,\gamma=\delta,\alpha \neq \gamma,\{\alpha,\gamma\}\in F\\
-\frac{u'_{\alpha\beta\gamma\delta}}{2}I,\hspace{0.5cm}&\alpha=\delta,\gamma=\beta,\alpha \neq \gamma,\{\alpha,\gamma\}\in F,
\end{cases}
\end{align}
while the two-body operators containing an odd number of modes in $F$ will be neglected.
We then map the fermionic Hamiltonians $H =\sum_{\alpha, \beta\neq 1s\sigma} t_{\alpha \beta} \, a'^\dag_{\alpha} a'_{\beta} +1/2  \sum_{\alpha, \beta, \gamma, \delta \neq 1s\sigma} u'_{\alpha \beta \gamma \delta}\, a'^\dag_{\alpha} a'^\dag_{\gamma} a'_{\delta} a'_{\beta}$ obtained in this way to our qubits.
The H$_2$ Hamiltonian is mapped first onto 4 qubits using a binary-tree mapping~\cite{BK2002S}. We order the $M$ spin-orbitals by listing first the $M/2$ spin-up ones and then the $M/2$ spin-down ones. When using the binary-tree mapping, this produces a qubit Hamiltonian diagonal in the second and fourth qubit, which has the total particle and spin $\bZ_2$ symmetries encoded in those qubits~\cite{TaperingS}. For the LiH and BeH$_2$ Hamiltonians we use the parity mapping, which has the two $\bZ_2$ symmetries encoded in the $M/2$-th and $M$-th mode, even if the total number of spin orbitals is not a power of 2, as in the case of H$_2$. We then assign to the Z Pauli operators of the $M/2$- and $M$-th qubits a value based on the total number of electrons $m$ in the system according to
\begin{equation}
\{Z_{M/2},Z_M\}=\begin{cases}
\{+1,+1\},\hspace{0.2cm}\mod(m,4)=0\\
\{\pm1,-1\},\hspace{0.2cm}\mod(m,4)=1\\
\{-1,+1\},\hspace{0.2cm}\mod(m,4)=2\\
\{\pm1,-1\},\hspace{0.2cm}\mod(m,4)=3,\\
\end{cases}
\end{equation}
The $+1$($-1$) on $Z_M$ for even(odd) $m$ implies an even(odd) total electron parity. The values $+1$, $-1$ and $\pm 1$ for $Z_{M/2}$ mean that the total number of electrons with spin-up in the ground state is even, odd, or there is an even/odd degeneracy, respectively. In the last case both $+1$ and $-1$ can be used equivalently for $Z_{M/2}$. The final qubit-tapered Hamiltonians consist of 4, 99 and 164 Pauli terms supported on 2, 4, 6 qubits, each having 2, 25 and 44 tensor product basis (TPB) sets (see Section~\ref{AppEnergyEst}) for H$_2$, LiH and BeH$_2$, respectively. We explicitly list the Hamiltonians at the bond distance in Table~\ref{HamTable}.

\section{Characterization of the Entanglers\label{EntSect}}

\begin{figure*}
\includegraphics[width=7in]{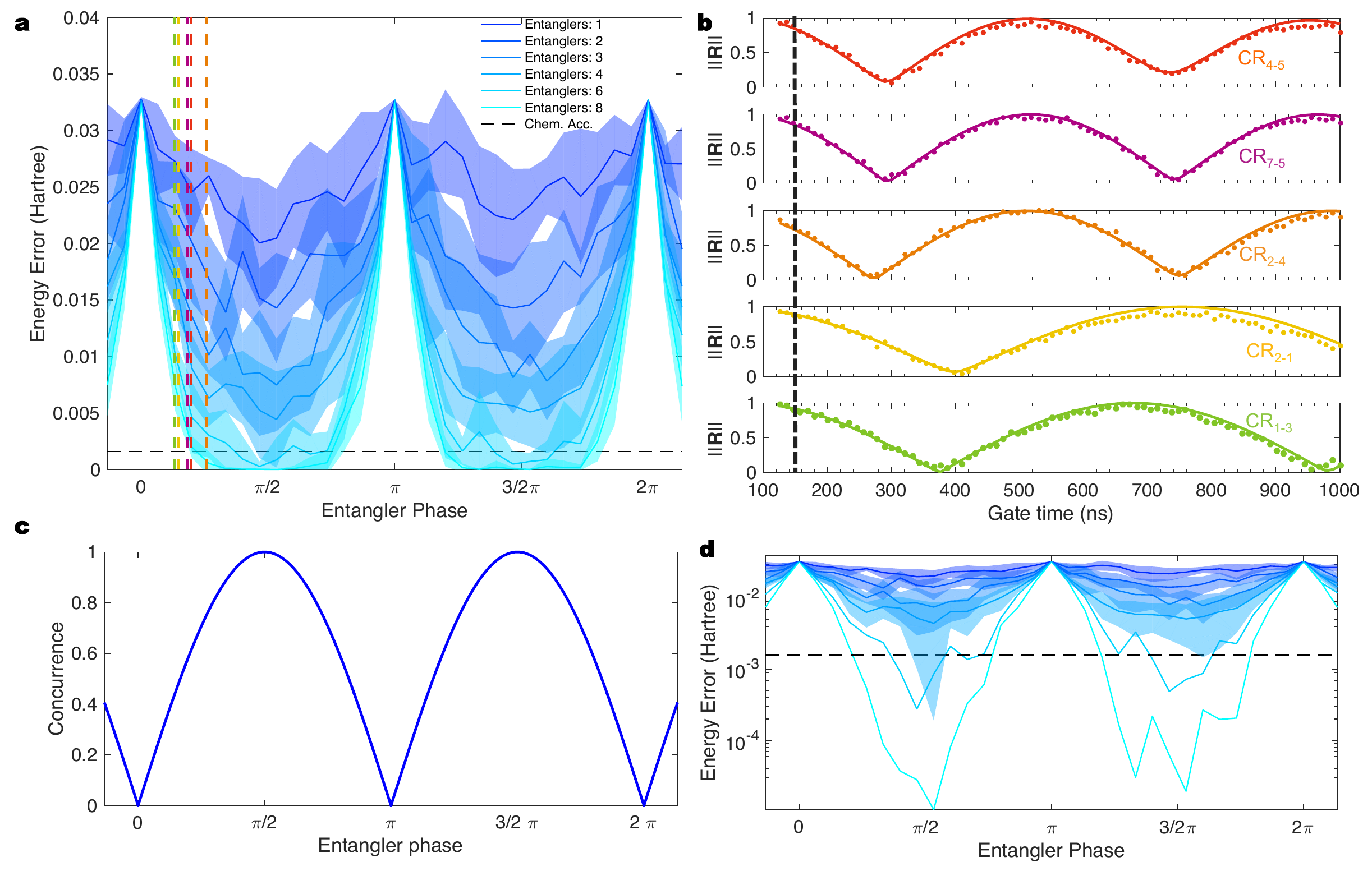}
\caption{\label{Supp_HamTomo} {\bf Dependence of energy error on entangler phase} {\bf a} Energy error of numerical optimizations, as a function of the phase of the entanglers, for different depths $d=1,2,3,4,6,8$. The energy error is averaged over 10 optimization runs, for each depth, with bands represent the standard deviation of the distribution. The dashed vertical lines indicate approximate gate phases of the individual CR gates for the gate time of 150~ns, including finite pulse ramping times. {\bf b} Norm of the Bloch vector $||\vec{R}||$ as a function of gate time for all the two-qubit entangling gates used in the experiment. The black dashed line corresponds to a gate time of 150~ns. The points where $||\vec{R}||=0$ indicate gate times of maximal entanglement. {\bf c} Concurrence v/s gate phase of a ZX gate, starting from the state $(\ket{10}+\ket{00})\sqrt{2}$. The energy error in {\bf a} is least around points of maximal concurrence. {\bf d} Energy error v/s entangler gate phase on a log linear scale. The dashed black line indicates chemical accuracy (0.0016 Hartree), showing that a critical depth $d=6$ is required to achieve such accuracy. Color scheme follows from {\bf a}.}
\end{figure*}

The entanglers in our hardware-efficient approach are collective gates composed of individual two-qubit gates on a convenient connectivity.
For our fixed frequency, multi-qubit architecture, a good choice of two-qubit entangling gate is the microwave-only cross resonance (CR) gate~\cite{Paraoanu2006S,Rigetti2010S,Chow2011S}. These gates constitute the entanglers $U_{\textrm{ENT}}$ in the trial state preparation and are implemented by driving a control qubit $Q_c$ with a microwave pulse that is resonant with a target qubit $Q_t$. With the addition of single qubit rotations, the CR gate can be used to construct a controlled NOT (CNOT), with fidelities exceeding $99\%$ for gate time $\sim$ 160 ns~\cite{Sheldon2016S}. In the hardware-efficient approach, however, tuning up a high-fidelity CNOT gate is not required, as long as entanglement is delivered with the CR drive. A simplistic model of the CR drive Hamiltonian is given by 
\begin{equation}
H_D\approx/\hbar\epsilon_{CR}(t)\Big(mIX-(J/\Delta)ZX+(\mu)ZI\Big)
\end{equation}
Here, $\epsilon_{CR}(t)$ is the CR drive amplitude, $m$ quantifies the strength of the classical cross-talk, $J$ is the strength of the qubit-qubit coupling, $\Delta$ is the frequency separation between the qubits, and $\mu$ corresponds to the drive induced Stark-shift. However, a more detailed study~\cite{Sheldon2016S} of the drive revealed additional terms, whose strengths are revealed by Hamiltonian tomography. For instance, in the CR$_{2-4}$ drive used in the experiment, these terms are $ZX: 1.04$ MHz, $ZY:$ 0.07 MHz,  $ZZ:$ 0.05 MHz , $IX:$ 0.68 MHz, $IY:$ 0.12 MHz, $IZ:$ 0.02 MHz. We measure the norm of the Bloch vector $||\vec{R}||$ discussed in~\cite{Sheldon2016S}, whose time evolution indicates points of maximal entanglement at $||\vec{R}||=0$; see Fig.~\ref{Supp_HamTomo}b. 

As discussed in the main text, the entangling gate phase could be an additional variational parameter for the optimization. However we show by numerical simulations that chemical accuracy ($\approx$ 0.0016 Hartree, the accuracy of the energy estimate required to predict the exponentially sensitive chemical reaction rates at room temperature to within an order of magnitude of the exact value) can be reached for a range of fixed gate phases around points of maximum concurrence. This is shown in Fig.~\ref{Supp_HamTomo}a,d which shows the error in the energy estimates from numerical optimization of the LiH Hamiltonian at bond distance, as a function of the gate phase of the two-qubit gates that compose the entanglers $U_\textrm{ENT}$.  For these simulations, we choose $ZX$ gates for $U_\textrm{ENT}$, using the same connectivity as the experiment (2-1, 1-3, 2-4 for the case of 4-qubit experiments). In order to isolate the effect of the entangling phase in the optimization, we do not consider a decoherence model and stochastic fluctuations in these simulations (as opposed to Fig.~3 and 4 in the main text), and set a high total number of energy evaluations to $5\times10^4$.  The results show plateaus of minimum energy errors, correlated with regions around points of maximal concurrence (Fig.~\ref{Supp_HamTomo}c) for the individual two-qubit gates. Instead of setting our gate times to points of maximal concurrence, we choose them such that the corresponding gate phases lie at the beginning of the minimal error plateaus, in order to minimize the effect of decoherence while delivering sufficient entanglement. For our chosen two-qubit gate time of 150 ns, we extrapolate the phases of all CR gates under the simple assumption of having a time independent $ZX$ Hamiltonian with finite pulse ramping times, and indicate them in Fig.~\ref{Supp_HamTomo}a. Also, CR drives for qubits on different buses are driven simultaneously, in order to reduce the time associated with state preparation.

\section{Energy estimation\label{AppEnergyEst}}

The update of the angles in our optimization routine is based on measurements of the expectation value of the Hamiltonian operator. These measurements are then used 
to build an approximation of the gradient of the energy landscape, which is in turn used to get a better update of the angles (see Section~\ref{SPSA}). The energy estimation at every $k$-th trial state of the optimization is a central 
part of the optimization algorithm, since its accuracy affects the final outcome of the optimization. 
Once mapped to qubits (see Section~\ref{HamDerivation}), every molecular Hamiltonian is expressed as a weighted sum of $T$ Pauli terms supported on $N$ qubits
\begin{equation}
\label{Hq}
H=\sum_{\alpha=1}^T h_\alpha P_\alpha,
\end{equation}
where each $P_\alpha\in \{X,Y,Z,I\}^{\otimes N}$ is a tensor product of single-qubit Pauli operators $X,Y,Z$ and the identity $I$, on $N$ qubits, with $h_\alpha$ being real coefficients.  
We are interested in estimating the mean energy $\bra{\Phi(\vec{\theta}_k)}H\ket{\Phi(\vec{\theta}_k)}\equiv\langle H\rangle_k$ for the $k$-th control updates (more specifically for two sets of angles close to $\vec\theta_k$, see Section~\ref{SPSA}). This can be done by averaging measurements outcomes from individual experiments, where one prepares the same initial state, applies the quantum gates parametrized by $\vec \theta_k$, and finally performs projective measurements on the individual qubits.
In the experiment we do not have access to direct measurements of the Hamiltonian operator $\langle H \rangle$ and its variance $\langle \Delta H^2\rangle=\langle H^2 -\langle H \rangle^2\rangle$. Instead, we sample the individual Pauli operators $P_\alpha$, estimating the mean values and variances $\langle P_\alpha \rangle$, $\langle \Delta P_\alpha^2\rangle=\langle P_\alpha^2-\langle P_\alpha\rangle^2\rangle$ from the measurements outcomes of the $\alpha$-th Pauli operator. The energy and Hamiltonian variance can then be obtained as 
\begin{align}
\label{MeanH}
\langle  H \rangle&=\sum_{\alpha=1}^T h_\alpha \langle P_\alpha \rangle,\\
\textrm{Var}[ H ]&=\sum_{\alpha=1}^T h_\alpha^2 \langle\Delta P_\alpha^2\rangle
\end{align}
Note that the variance on the mean energy $\textrm{Var}[{ H }]$ is different from ${\langle \Delta H^2 \rangle}$, since we are sampling the individual Pauli terms separately: for example, eigenstates of $H$ will have ${\langle \Delta H^2 \rangle}=0$, but a finite $\textrm{Var}[{ H }]\neq0$.
The error on the mean energy $\langle H\rangle$ after taking $S$ samples for each Pauli operator is 
\begin{align}
\label{ErrMean}
\epsilon=\sqrt{\frac{\textrm{Var}[ H]}{S}}\leq \sqrt{\frac{T|h_\textrm{max}^2|}{S}}
\end{align}
where $h_{\textrm{max}}=\max_{\alpha} |h_\alpha|$ is the absolute value of the largest Pauli coefficient. Since sampling $S$ times for a large number of trial states and Pauli operators comes with significant time overhead, one can instead use the same state preparations to measure different Pauli operators. This approach was considered in~\cite{mcclean2016theoryS} for commuting operators. Here we use a stronger condition on grouping different Pauli terms, based on improving time efficiency. 
We fist briefly describe how we sample an individual Pauli operator. The individual Pauli operators are measured by correlating measurement outcomes of single-qubit dispersive readouts in the $Z$ basis, which can be done simultaneously since each qubit is provided with an individual readout resonator. In case a target multi-qubit Pauli operator contains non diagonal single-qubit Pauli operator, single-qubit rotations (post-rotations) are performed before the measurement in the $Z$ basis. Specifically, a $-\pi/2$($\pi/2$) rotation along the $X$($Y$) axis to measure a Y(X) single-qubit Pauli operator. 

\subsection{Grouping Pauli Operators}
To minimize sampling overheads, we group the $T$ Pauli operators $P_\alpha$ in $A$ sets $s_1,s_2,...s_A$, which have terms that are diagonal in the same tensor product basis. The post-rotations required to measure all the Pauli terms in a given TPB set are the same, and a unique state preparation can be used to sample all the Pauli operators in the same set.  
\begin{figure*}
\includegraphics[width=7in]{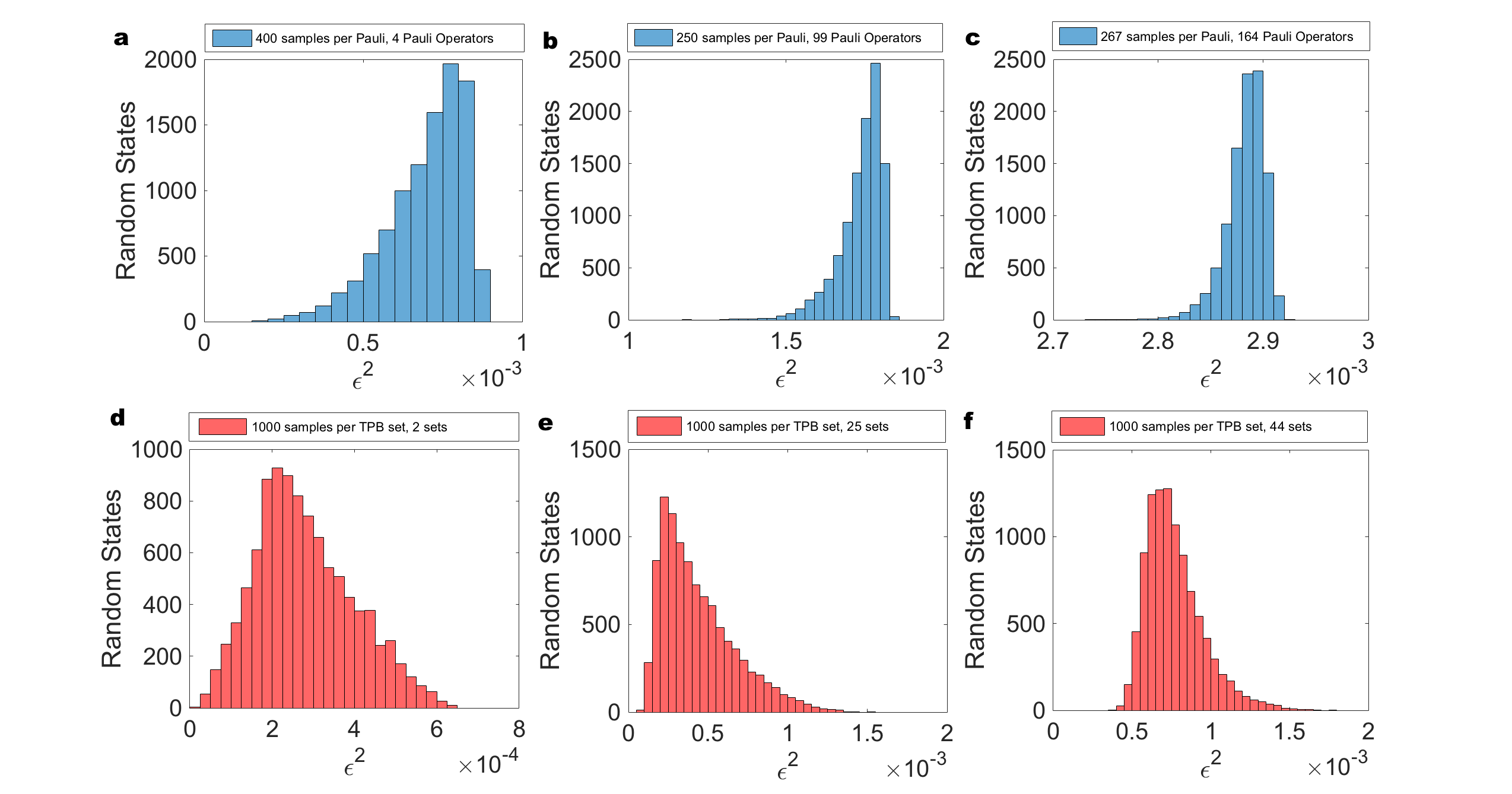}
\caption{\label{SupplVarPlot} {\bf Energy variance} Numerical computation of the variance of the mean energy $\epsilon^2$, as in Eq.~(\ref{VarEst}), with $S=10^3$ samples, for the molecular Hamiltonians of $\textrm{H}_2$ ({\bf a, d}), $\textrm{LiH}$ ({\bf b, e}) and $\textrm{BeH}_2$ ({\bf c, f}) at their bond interatomic distances (see Table~\ref{HamTable}). The variances are computed sampling each Pauli operator $P_\alpha$ in $H$ of Eq.~(\ref{Hq}) individually ( {\bf a, b, c} ) and grouping them in TPB sets ( {\bf d, e, f} ), keeping the total number of samples the same.}
\end{figure*}
By doing so, however, covariance effects in the same TPB set contribute to the 
variance of the total Hamiltonian,
\begin{equation}
\label{VarCov}
\textrm{Var}^G[{ H }]=\sum_{i=1}^A \sum_{\alpha,\beta\in s_i} h_\alpha h_\beta\langle(P_\alpha-\langle P_\alpha \rangle)(P_\beta-\langle P_\beta \rangle)\rangle\leq h_\textrm{max}^2 (T+As_{\textrm{max}}^2),
\end{equation}
where $s_{\textrm{max}}=\max_{i} |s_i|$ is the number of elements in the largest TPB set. 
Keeping the same total number of measurements $TS$ as in Eq.~(\ref{ErrMean}), the error on the mean in this case is given by 
\begin{align}
\label{ErrMeanGroup}
\epsilon=\sqrt{\frac{\textrm{Var}^G[ H]}{S}}\leq \sqrt{\frac{ Ah_\textrm{max}^2 (T+As_{\textrm{max}}^2)}{TS}},
\end{align}
which can be compared to the case in which one samples the single Pauli terms individually, Eq.~(\ref{ErrMean}). The error contribution from the covariance (which can be positive or negative) has to be traded off against the use of less samples from grouping.
The quantities in Eqs.~(\ref{MeanH}) and (\ref{VarCov}) can be estimated in the experiment and in the numerical simulations as 
\begin{align}
\label{MeanEst}
\widehat{\langle P_\alpha \rangle}&=\frac{1}{S}\sum_{i=1}^S X_{i,\alpha},\\
\label{VarEst}
\widehat{\textrm{Var}^G[{ H }]}&=\sum_{i=1}^A \sum_{\alpha,\beta\in s_i} h_\alpha h_\beta{\mathrm{cov}}(\widehat{\langle P_\alpha\rangle},\widehat{ \langle P_\beta\rangle}),
\end{align}
where we have defined the outcome of the $i$-th measurement on the $\alpha$-th Pauli term as $X_{i,\alpha}$. The covariance matrix element is defined after $S$ measurements as 
\begin{equation}
\label{CovDef}
{\mathrm{cov}}(\widehat{ \langle P_\alpha\rangle},\widehat{ \langle P_\beta \rangle})=\frac{1}{S-1}\sum_{i=1}^S (X_{i,\alpha}-\widehat{ \langle P_\alpha\rangle}_k)(X_{i,\beta}-\widehat{ \langle P_\beta\rangle}).
\end{equation}
To evaluate whether grouping into TPB sets is convenient for the molecular Hamiltonians considered in this work, we perform numerical sampling experiments, shown in Fig.~\ref{SupplVarPlot}, using the Hamiltonians in Table~\ref{HamTable}. The variance of the mean energy is numerically sampled on $10^4$ random states. In the ``TPB sets'' simulations (red histograms), the set of post-rotations associated to each TPB set if found by union of the set of post-rotations necessary to sample each Pauli in a given TPB set: for example, for the third TPB set of BeH$_2$ in Table~\ref{HamTable} we have the post-rotations associated to ZZXXZX. Then, for each random state, a sample of $S=10^3$ measurement outcomes are drawn for every TPB set. The total number of measurement is therefore $AS$. These measurements are then used to obtain the mean value and covariance for each Pauli operator in the TPB set. The variance of the mean total energy is then obtained as in Eq.~(\ref{VarCov}). In the ``No-TPB sets'' simulations (blue histograms), the same measurements are drawn independently for each Pauli operator, with a number of samples per Pauli term $SA/T$, in order to keep the total number of samples in the TPB and No-TBP simulations the same. The results show the advantage of grouping into TPB sets for all the molecular Hamiltonians considered.

\subsection{Assignement Errors}
An important aspect to take into account when sampling is the presence of assignment errors at the qubit readout. A qubit-independent assignement error can be modeled by a deformation $\hat{\Pi}_0,\hat{\Pi}_1$, of the ideal projectors $\Pi_0$, $\Pi_1$ on the $\ket{0},\ket{1}$ states for the qubit, 
\begin{align}
\hat{\Pi}_0&=(1-\eta_0+\eta_1)\Pi_0+(1-\eta_0-\eta_1)\Pi_1=(1-\eta_0)\textrm{I}+\eta_1 Z\nonumber\\
\hat{\Pi}_1&=(\eta_0-\eta_1)\Pi_0+(\eta_0+\eta_1)\Pi_1=\eta_0\textrm{I}-\eta_1 Z,\label{DefProj}
\end{align}
via the two parameters $\eta_0,\eta_1$ (note that in the absence of errors $\eta_0=\eta_1=1/2$), such that $\hat{\Pi}_0+\hat{\Pi}_1=\textrm{I}$. With these definitions, the assignment error of reading a qubit in $\ket{1}(\ket{0})$ when it is in $\ket{0}(\ket{1})$ is given by $1-\eta_0-\eta_1$, or ($\eta_0-\eta_1$). The measured readout assignement error, averaged on preparations of $\ket{0}$ and $\ket{1}$  in Table~\ref{table:device_parameter}, can be expressed with the parametrization considered as $\epsilon_r=1/2-\eta_1$. The projectors in Eqs.~(\ref{DefProj}) define an effective deformed $\hat{Z}$ operator, related to the ideal one $Z$ via
\be
\hat{Z}=\hat{\Pi}_0-\hat{\Pi}_1, \hspace{1cm}Z=\frac{\hat{Z}-(1-2\eta_0)\textrm{I}}{2\eta_1}.
\ee  
Note that the measured value $\langle\widehat{ Z }\rangle$ is affected by the contrast factor $2\eta_1$, and shifted by the amount $1-2 \eta_0$. Generalizing  this to a Pauli operator with weight $w$, one has that 
\begin{align}
Z^{\otimes w}&\propto \frac{\hat{Z}^{\otimes w}}{(2\eta_1)^w},\label{ExpNoise}
\end{align}
revealing an exponential loss in contrast in the weight $w$. When addressing larger systems, it will then be important to use the binary tree encoding~\cite{BK2002S}, for its logarithmic scaling in locality with the system size, to combat the exponential scaling in~(\ref{ExpNoise}). Note that the error model in Eq.~(\ref{DefProj}) only takes into account independent readout errors, while in general correlated readout errors may happen. In our experiments we take into account assignement errors by running readout calibrations before sampling for every update of the angles $\vec\theta$, and then correcting our sampling outcome with the calibrations.

\section{Optimization using a simultaneous perturbation method\label{SPSA}}

The energy $\bra{\Phi(\vec{\theta}_k)}H\ket{\Phi(\vec{\theta}_k)}\equiv\langle H\rangle_k$ discussed in Section~\ref{AppEnergyEst}, which needs to be evaluated before every update of the angles $\vec\theta$, has a number of parameters $p=N(3d-1)$ that grows linearly with the depth of the circuit d and the number of qubits $N$. 
As the number of parameters increases the classical optimization component of the algorithm comes with increasing overheads. The accuracy of the optimization may also be significantly lowered by the presence of energy fluctuations at the $k$-th step $\epsilon_k$.  Furthermore, on real quantum hardware, there are time overheads associated with loading of pulse waveforms on the electronics, resonator and qubit reset, and repeated sampling of the qubit readout. Ideally, one would like to use an optimizer robust to statistical fluctuations, that uses the least number of energy measurements per iteration. The simultaneous perturbation stochastic approximation (SPSA) algorithm, introduced in~\cite{Spall1992S}, is a gradient-descent method that gives a level of accuracy in the optimization of the cost function that is comparable with finite-difference gradient approximations, while saving an order $\cO (p)$ of cost function evaluations. It has been recently used in the context of quantum control and quantum tomography~\cite{Ferrie2014S,Ferrie2015S,Chapman16S}. 

\begin{figure*}
\includegraphics[width=6in]{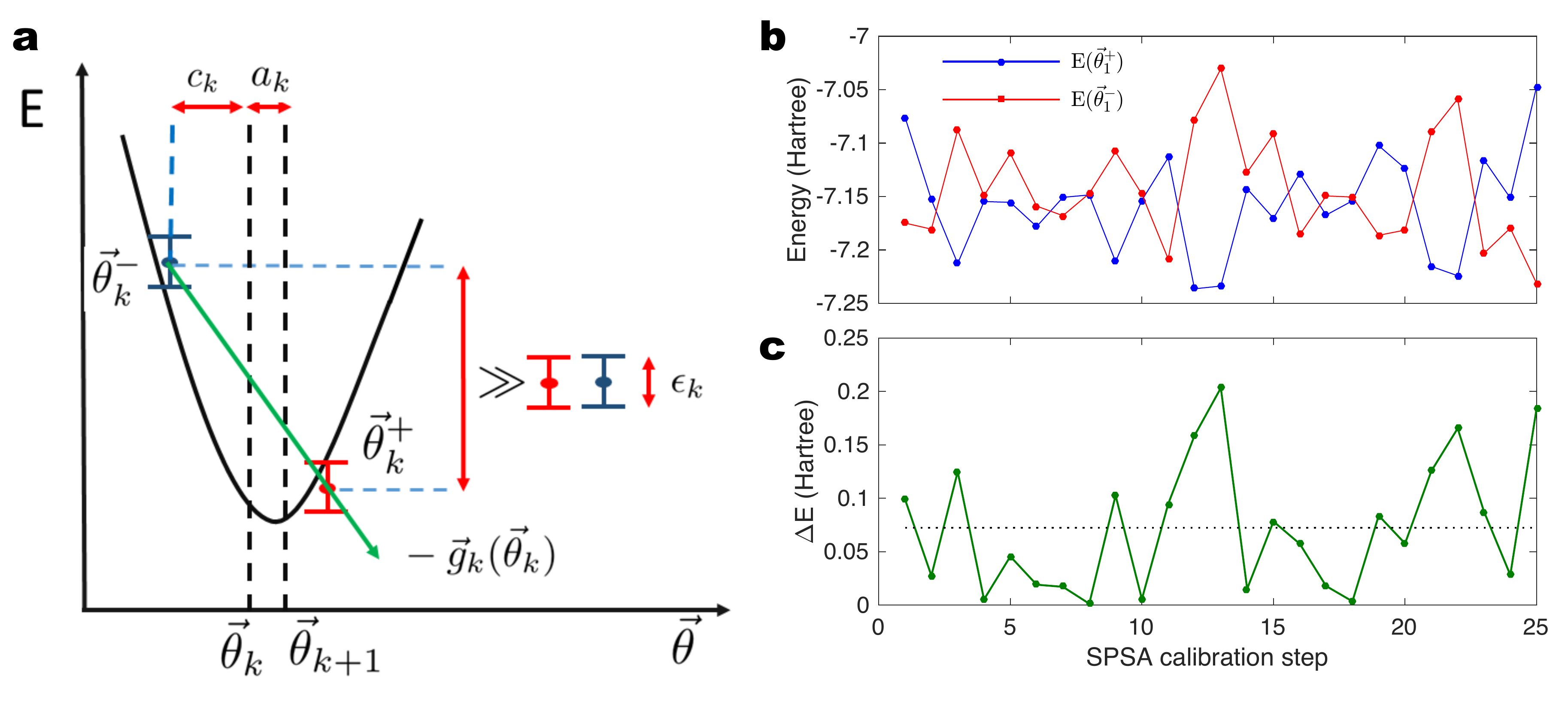}
\caption{\label{Supp_SPSA1search} {\bf Calibration of the classical optimizer }{\bf a} Good gradient approximations $\vec{g}_k(\vec\theta_k)$ are obtained if the energy difference $|\langle\Phi(\vec\theta^+_k)|H|\Phi(\vec\theta^+_k)\rangle-\langle\Phi(\vec\theta^-_k)|H|\Phi(\vec\theta^-_k)\rangle|$ is larger than the stochastic fluctuations on the energy $\epsilon_k$. The parameter $c$ in Eq.~(\ref{alphagamma}) is heuristically chosen to meet this condition. {\bf b} The parameter $a$ in Eq.~(\ref{alphagamma}) is calibrated by measuring 25 times the energies $E(\vec\theta_1^\pm)=\langle\Phi(\vec\theta^\pm_1)|H|\Phi(\vec \theta^\pm_1)\rangle$, measured here for the LiH molecule at the bond distances, from the starting angles $\vec\theta_1$, for different random gradients approximations. {\bf c} The energy difference $\Delta E=|\langle\Phi({\vec \theta^+_1})|H|\Phi({\vec \theta^+_1})\rangle-\langle\Phi({\vec \theta^-_1})|H|\Phi({\vec \theta^-_1})\rangle|$ is measured for each random instance of the gradient (solid green line), averaged (black dotted line), and then used to calibrate the parameter $a$, according to Eq.~(\ref{avggrad}).}
\end{figure*}

In the SPSA approach, for every step $k$ of the optimization, we sample from $p$ symmetrical Bernoulli distributions (coin flips) $\vec\Delta_k$, and use preassigned elements from two sequences converging to zero, $c_k$ and $a_k$. The gradient at ${\vec \theta}_k$ is approximated using energy evaluations at ${\vec \theta}^\pm_k={\vec \theta}_k\pm c_k{\vec \Delta}_k$, and is constructed as
\begin{equation}
\label{SPSAGrad}
{\vec g}_k({\vec \theta_k})=\frac{\langle\Phi({\vec \theta^+_k})|H|\Phi({\vec \theta^+_k})\rangle-\langle\Phi({\vec \theta^-_k})|H|\Phi({\vec \theta^-_k})\rangle}{2c_k}{\vec \Delta}_k,
\end{equation} 
as illustrated in Fig.~\ref{Supp_SPSA1search}a. Note that this gradient approximation only requires two estimations of the energy, regardless of the number $p$ of variables in $\vec\theta$. The controls are then updated as 
\begin{equation}
\label{CUpdate}
{\vec \theta}_{k+1}={\vec \theta}_k-a_k {\vec g}_k(\vec\theta_k).
\end{equation}
The convergence of $\theta_k$ to the optimal solution ${\vec \theta}^*$ can be proven even in the presence of stochastic fluctuations, if the starting point is in the domain of the attraction of the problem~\cite{Spall1992S}, . Convergence remains an open issue if the starting point for the controls is not in a domain of attraction. In this case strategies like multiple competing starting points can be adopted~\cite{Wecker2015S}.  
The sequences $c_k,a_k$ can be chosen as
\begin{align}
c_k&=\frac{c}{k^\gamma},\nonumber\\
\label{alphagamma}a_k&=\frac{a}{k^\alpha}.
\end{align}
We pick the parameters $\alpha, \gamma$ optimally at $\{\alpha,\gamma\}=\{0.602,0.101\}$~\cite{Spall1998S}, ensuring the smoothest descent along the approximate gradients defined in Eq.~(\ref{CUpdate}). 
We then tune the value of $c$ to adjust the robustness of the gradient evaluation with respect to the magnitude of the energy fluctuations. In fact, large fluctuations of the energy require gradient evaluations with large $c_k$ (\ref{SPSAGrad}), so that the fluctuations do not substantially affect the gradient approximation. This condition is valid in the regime 
\begin{equation}
\label{GradRegime}
|\langle\Phi(\vec\theta^+_k)|H|\Phi(\vec\theta^+_k)\rangle-\langle\Phi(\vec\theta^-_k)|H|\Phi(\vec\theta^-_k)\rangle| \gg \epsilon_k,
\end{equation}
depicted visually in Fig.~\ref{Supp_SPSA1search}a. Keeping these considerations in mind, we have used $c=10^{-1}$ to ensure robustness in all the experiments and in the realistic simulations that include decoherence noise and energy fluctuations, while the smaller $c=10^{-2}$ factor is used in the numerical optimizations where the energy is evaluated without fluctuations. 
The parameter $a$ is then calibrated experimentally in order to achieve a reasonable angle update on the first step of the optimization, which we chose to be $|\theta^{(i)}_{2}-\theta^{(i)}_{1}|=2\pi/10$, for all the angles $i=1,2,...p$. To achieve this, we use an inverse formula based on Eq.~(\ref{CUpdate}),
\begin{equation}
\label{avggrad}
a= \frac{2\pi }{5}\frac{c}{\Big\langle|\langle\Phi({\vec \theta^+_1})|H|\Phi({\vec \theta^+_1})\rangle-\langle\Phi({\vec \theta^-_1})|H|\Phi({\vec \theta^-_1})\rangle|\Big\rangle_{\vec\Delta_1}},
\end{equation}
where the notation $\Big\langle\Big\rangle_{\vec\Delta_1}$ indicates an average over different samples from the distribution $\vec\Delta_1$ that generates the first gradient approximation. In fact, by averaging along different directions, we can measure the average slope of the functional landscape of $\langle\Phi(\vec\theta)|H|\Phi(\vec\theta)\rangle$ in the vicinity of the starting point $\vec\theta_1$, and calibrate the experiment accordingly. In the experiment and in the numerics the average $\Big\langle\Big\rangle_{\vec\Delta_1}$ is realized over 25 random gradient directions. The gradient averaging is shown for the optimization of the LiH Hamiltonian at bond distance with a $d=1$ circuit, in Fig.~\ref{Supp_SPSA1search}b,c.

Note that along the optimization we do not measure the value of the energy for the $k$-th optimized angles $\langle\Phi(\vec\theta_k)|H|\Phi(\vec\theta_k)\rangle$, instead we only measure and report the values $\langle\Phi(\vec\theta^+_k)|H|\Phi(\vec\theta^+_k)\rangle$ and $\langle\Phi(\vec\theta^-_k)|H|\Phi(\vec\theta^-_k)\rangle$, which serve to generate a new gradient approximation. The underlying optimized angles $\vec\theta_k$ are only measured at the end of the optimization, averaging over the last 25 $\vec\theta^+_k$ and 25 $\vec\theta_k^-$, to further minimize stochastic fluctuations effect. Furthermore, this last average is done with $10^5$ samples, as opposed to the $10^3$ samples used to generate $\vec\theta_k^+$ and $\vec\theta_k^-$ during the optimization, in order to reduce the error on the measurement.

\section{Numerical simulations and scaling of resources\label{Scaling}}

In this Section we first describe the numerical simulations used in Fig.~3 and Fig.~4, which include decoherence effects and stochastic fluctuations on the energy evaluation. We then show numerical results that indicate the scaling of the optimization outcome with the depth of the trial state preparation circuit, the number of angle updates considered in the optimization, and the sampling statistics. We estimate the resources necessary to achieve chemical accuracy for the three molecules considered. Last, we show the interplay between circuit depth and decoherence affecting the quantum circuit, using a depolarizing noise model. 

\subsection{Numerical model of the experiment}
In the numerical simulations in Fig.~3, Fig.~4 and Fig.~\ref{ExpDepth}, we have used entanglers made up of $ZX$ two-qubit entangling gates, with a phase of $\pi/4$, and with additional terms $ZY$, $ZZ$, $IX$, $IY$, and $IZ$, whose relative phases are chosen according to the measurement reported in Section~\ref{EntSect} for CR$_{2-4}$. We use the same connectivity as in the experiment, with entangling gates between qubits $1-2$, $2-4$ and $1-3$ in the $4$-qubit simulations (LiH and quantum magnetism model) and gates between qubits $1-2$, $2-4$, $1-3$, $4-5$ and $5-6$ in the $6$-qubit simulations (BeH$_2$). The initial $Z$ angles are distributed normally around zero according to $\mathcal{N}(0,1)$, and the $X$ angles set to $\pi/2$. 

The effect of decoherence is taken into account by adding amplitude damping ($E^a_0(\tau),E^a_1(\tau)$) and dephasing ($E^d_0(\tau),E^d_1(\tau)$) channels acting on the system density matrix $\rho\rightarrow E^a_0(\tau) \rho E^{a\dag}_0(\tau)+E^a_1(\tau) \rho E^{a\dag}_1(\tau),\rho\rightarrow E^d_0(\tau) \rho E^{d\dag}_0(\tau)+E^d_1(\tau) \rho E^{d\dag}_1(\tau)$, for all the qubits, after each round of Euler gates and entanglers, respectively.  The strength of the channels is set by the experimental coherence times and the length of the gates,
\begin{align}
E^a_0(\tau)=\begin{bmatrix}
    1      & 0 \\
    0    & \sqrt{e^{-\tau/T_1}} 
\end{bmatrix},
E^a_1(\tau)=\begin{bmatrix}
    0      & \sqrt{1-e^{-\tau/T_1}} \\
    0    & 0 
\end{bmatrix}\\
\nonumber\\
E^d_0(\tau)=\begin{bmatrix}
    1      & 0 \\
    0    & e^{-\tau/T_\phi} 
\end{bmatrix},
E^d_1(\tau)=\begin{bmatrix}
    0      &0 \\
    0    &  \sqrt{1-e^{-2\tau/T_\phi}}
\end{bmatrix}.
\end{align}
Here the time $\tau$ alternates between the duration of each single qubit gate sequence or entangler step, and the pure dephasing time is defined as $T_\phi=2T_2^*T1/(2T_1-T_2^*)$, see 
Table~\ref{table:device_parameter} for measured values on each qubit. 
In the $\textrm{H}_2$ simulations, since we use the most coherent qubits on the chip, we parametrize the noise channels considering $T_1=T_2^*=40~\mu$s and set the length of $U_{\textrm{ENT}}$ to $150$~ns, while for the 4 and 6-qubit simulations we use typical coherence values for the qubits of $T_1=30~\mu$s, $T_2^*=20~\mu$s and a duration for $U_{\textrm{ENT}}$ of $450$~ns. Note that the duration for both 4 and 6-qubit entanglers is set to be the same because the two-qubit gates CR$_{2-1}$, CR$_{4-5}$ and CR$_{1-3}$, CR$_{6-5}$ are done in parallel, see Fig.~1c in the main text.
To simulate the effect of finite sampling in the experiment, we first compute an average value of the standard deviation of the energy by sampling $10^3$ times on $100$ random states, as described in Section~\ref{AppEnergyEst}. Then we add a normal-distributed error to each energy evaluation along the optimization, with the standard mean deviation computed previously on random states. On average, this will account for the energy fluctuations at the $k$-th step of the optimization. We fix the total number of angle updates to 250. For the final energy estimate, we average over the last 25 control updates, to mitigate the effect of stochastic fluctuation in the optimization. For every interatomic distance (for every $J/B$ ratio in the case of Fig.~4), we show the outcome of 100 numerical simulations, in the form of a density plot, in Fig.~3 (Fig.~4) in the main text.

\begin{figure*}
\includegraphics[width=6in]{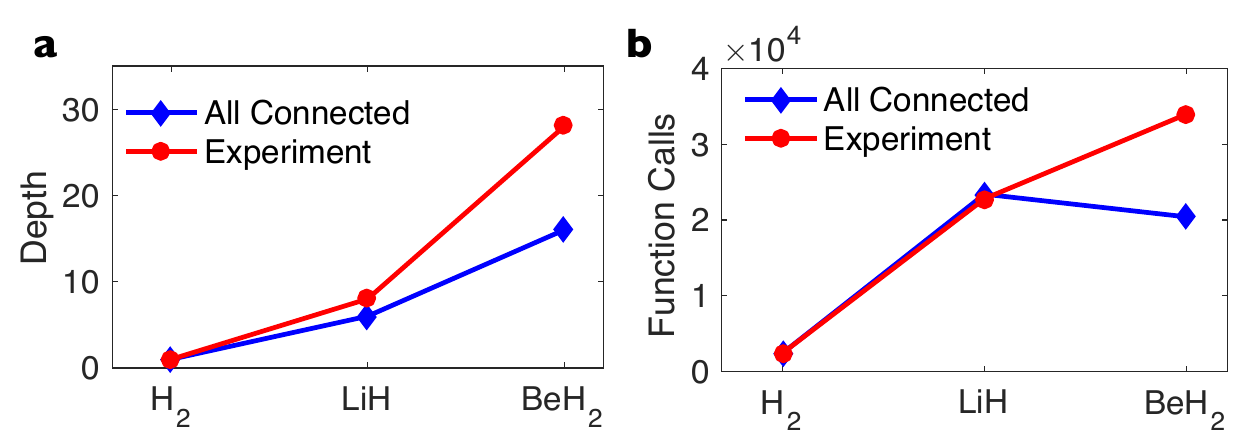}
\caption{\label{Supp_ChemAcc} {\bf Scaling of resources to reach chemical accuracy}. {\bf a} The critical depth required for reaching chemical accuracy for the 3 molecules discussed in the paper, using an all-to-all qubit connectivity (blue) and the experimental qubit connectivity (red). {\bf b} The number of function calls for reaching chemical accuracy for the 3 molecules at their respective critical depths from {\bf a}. Each data point in both plots is obtained by averaging over 10 optimization runs.}
\end{figure*}

\subsection{Scaling of resources: depth, function calls, sampling}

In order to estimate resources required to reach chemical accuracy (i.e. an energy error of approximately $0.0016 $ Hartree), we consider molecular Hamiltonians at the bond distance for H$_2$, LiH and BeH$_2$ (see Table~\ref{HamTable}), and declare convergence when the best energy estimate is close to the exact solution up to chemical accuracy. We assume that the resources required to reach chemical accuracy at the bond distance are comparable with the ones for any other interatomic distance, ensuring chemical accuracy also for the dissociation energy (defined as the molecular energy difference at the bond length and in the limit of infinite interatomic energy). In these simulations for determining the scaling of the resources, we consider ideal $ZZ$ entangling gates with a phase of $\pi/2$. Note that any two qubit interaction can be mapped to a ZZ one via local rotations (i.e. our Euler angles). We use only the last two single-qubit rotation for each step, since Z rotations commute with the ZZ entangling gates, and consider two different topologies for the qubit connectivity: in addition to the experimental connecivity, we consider an ``all connected'' connectivity, where the entanglers $U_{\textrm{ENT}}$ are composed of ZZ gates among all the qubit pairs in the system.

For the simulations outcomes plotted in Fig.~\ref{Supp_ChemAcc}a, we set a maximal number of function calls to $5\times 10^4$ (i.e. evaluations of the energy as described in Section~\ref{SPSA}), ensuring convergence of the optimization beyond chemical accuracy for all the simulations considered. We start by not taking into account decoherence and stochastic fluctuations, run 10 optimizations for increasing circuit depths, average the final optimized energies, and report the shortest depth that has an average energy converged within chemical accuracy. Chemical accuracy is reached for depths $d=1,8,28$ for the experimental connectivity,  and $d=1,6,16$ for the all connected case, for H$_2$, LiH and BeH$_2$, respectively.
Having computed the shortest circuit depth for each molecule and connectivity, we now keep the circuit depth fixed and run optimizations, keeping track of the number of trial states sufficient to achieve chemical accuracy. We average the number of trial states obtained for 10 separate optimizations. The results are plotted in Fig.~\ref{Supp_ChemAcc}b. Approximately $2\times 10^3$ function calls ($10^3$ angle updates) are sufficient for reaching chemical accuracy on H$_2$, $2\times 10^4$ for LiH$_2$ both for the all-connected and experiment connectivity, $2\times 10^4$ for BeH$_2$ in the all-connected case and approximately $3\times10^4$ for the experiment connectivity. 

We finally estimate the number of samples $S$ required to reach chemical accuracy. We start by computing an average standard deviation $\epsilon_A$ for the energy on $10^2$ random states, considering $S=10^3$ samples, see Section~\ref{AppEnergyEst}. Then we add the averaged deviation to the energies evaluated at the $k$-th step of the optimization. Then, we extrapolate standard deviations at higher samplings $S$, via $\epsilon_A\rightarrow\epsilon_A\sqrt{10^3/S}$. Using the depths indicated in Fig.~\ref{Supp_ChemAcc}a, we find that chemical accuracy is reached for all the three molecules when the number of samples is $S\approx 10^6$, i.e. approximately when all the energies in the optimization are evaluated at chemical accuracy. This can be understood by using values for the standard deviations of the mean energies as in Fig.~\ref{SupplVarPlot}, computed at $10^3$ samples, and extrapolating to $10^6$ samples.  
These results indicate a scaling of the resources with the problem size which is not very dramatic. If we set aside decoherence effects, both number of function calls and sampling could be increased in the near future by rapid reset protocols of the qubits~\cite{Riste2012S,ResetS,Bultnick2016S}.

\subsection{Scaling of resources: decoherence}

\begin{figure*}
\includegraphics[width=5in]{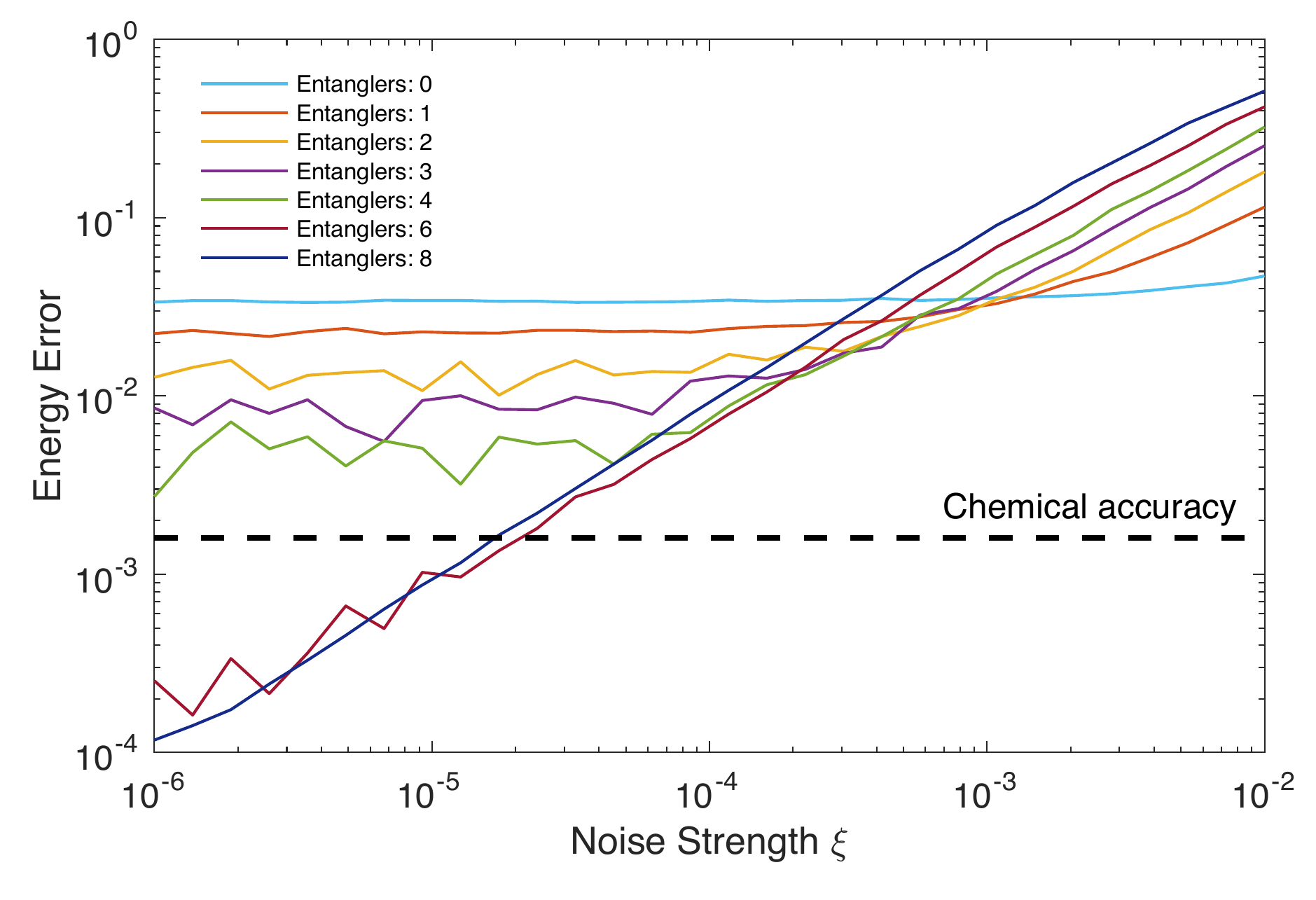}
\caption{\label{Supp_LiHErrorScaling}  {\bf Scaling of energy error with noise strength} Error in the energy estimate for the 4-qubit LiH Hamiltonian at its bond length, for different depolarizing noise strengths of the model in Eq.~(\ref{Depo}), for different circuit depths used for trial state preparation, after $5\times10^4$ function calls. Each data point is obtained by averaging over 10 optimization runs. The black dashed line indicates the energy error for chemical accuracy.}
\end{figure*}

In order to address the behavior of the optimization versus decoherence effects, we run numerical simulations that include a depolarizing noise model following each gate. We consider one-qubit and two-qubit depolarizing channels acting on the system density matrix $\rho$ as 
\begin{align}
\label{Depo}
\rho&\rightarrow(1-\xi)\rho +\frac{\xi}{3} \sum_{i=1,2,3} \sigma^i\rho \sigma^i,\nonumber\\
\rho&\rightarrow(1-\xi)\rho +\frac{\xi}{15} \sum_{\substack{\{i,j\}=\{0,1,2,3\}\\\{i,j\}\neq\{0,0\}}} \sigma^j_l\sigma^i_m\rho \sigma^i_m\sigma^j_l,
\end{align}
where $\sigma^1=X,\sigma^2=Y,\sigma^3=Z,\sigma^0=\textrm{I}$. The single-qubit depolarizing channels act on every qubit after the Euler rotations, while the two-qubit channels act on every qubit pair $\{l,m\}$ considered in a given connectivity. We run noisy optimizations for the LiH Hamiltonian at the bond distance, for different number of entanglers and noise strengths, for a maximum of $5\times10^4$ function calls. The results are shown in Fig.~\ref{Supp_LiHErrorScaling}, averaged on 10 different optimizations. There is a clear interplay between the number of entanglers and the noise strength. For low noise rates $\xi$, higher depths give better results, while as $\epsilon$ increases lower depths perform better. Chemical accuracy is reached for noise rates of $\approx 10^{-5}$, for 6 and 8 entanglers. Such low noise rates emphasize that it will be important in the near future to explore error mitigation methods for short depth quantum circuits~\cite{MitigationS,Mitigation2S,Mitigation3S}.

\begin{figure*}
\includegraphics[width=6in]{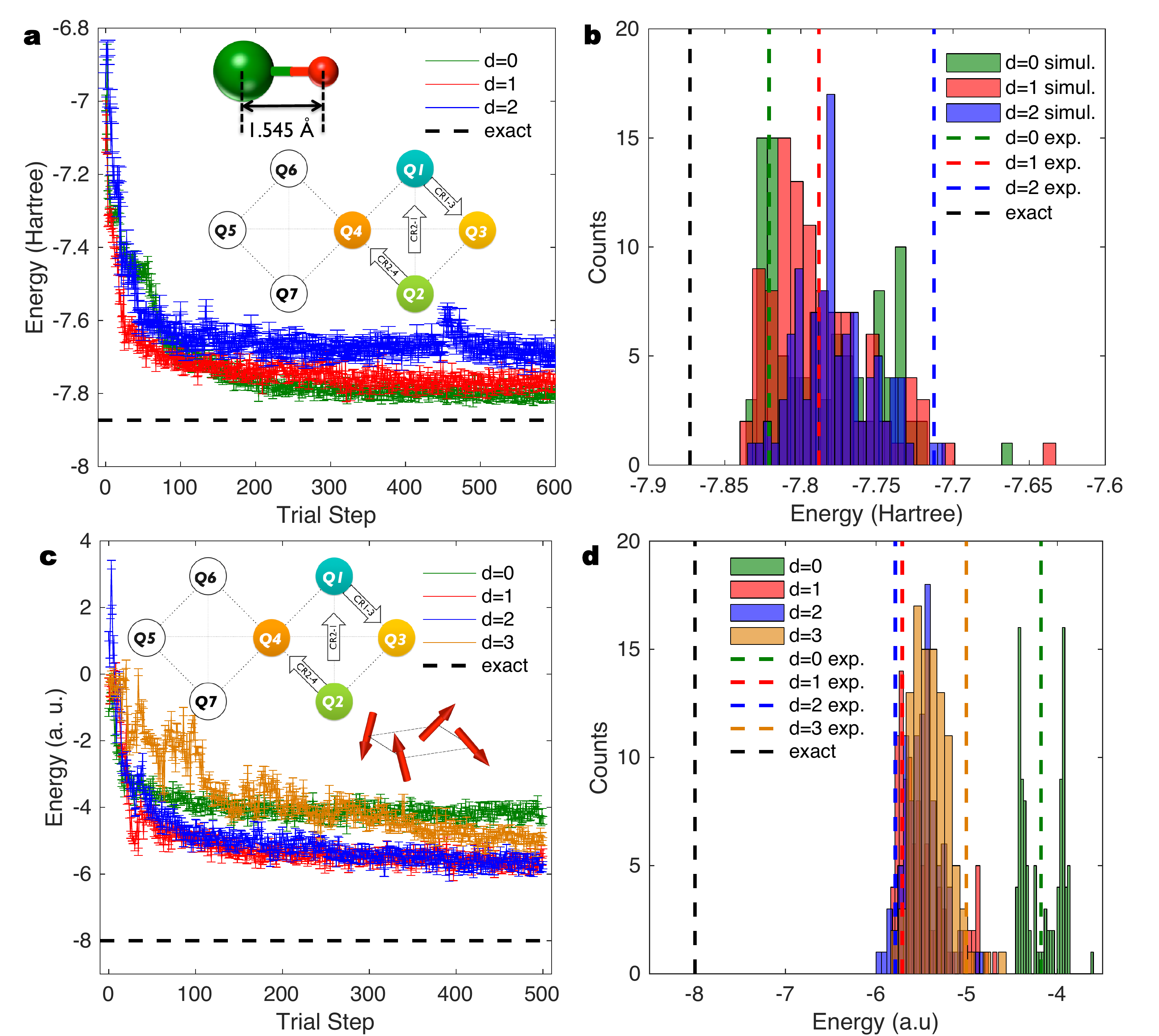}
\caption{{\bf Experimental optimization for different depths: LiH Hamiltonian at bond distance and 4-qubit Heisenberg model} {\bf a} Experimental optimization of  the 4-qubit LiH Hamiltonian at bond distance, using depth $d=$ 0 (green), 1 (red) 2 (blue) circuits for trial state preparation. The exact energy is indicated by the black dashed line. Bottom inset describes the qubits and the cross resonance gates that constitute $U_{\textrm{ENT}}$, for this experiment. {\bf b} Histograms of outcomes from 100 numerical simulations that account for decoherence and finite sampling effects show significant overlap for depth $d=$ 0 (green), 1 (red), 2 (blue) circuits. The black dashed line indicates the exact energy and the green, red and blue dashed lines are the results from the single experimental runs of {\bf a}, for  $d=$ 0, 1 and 2 circuits respectively.\label{ExpDepth} {\bf c} Experimental optimization of  the 4-qubit Heisenberg Hamiltonian for $J/B=1$, using depth $d=$ 0 (green), 1 (red), 2 (blue), 3 (orange) circuits for trial state preparation. The exact energy is indicated by the black dashed line. {\bf d} Histograms of outcomes from 100 numerical simulations that account for decoherence and finite sampling effects show significant improvement over depth $d=$ 0 circuits with $d=$ 1(red), 2 (blue), 3 (orange) circuits. The black dashed line indicates the exact energy and the green, red, blue and orange dashed lines are the results from the single experimental runs of {\bf c}, for  $d=$ 0, 1, 2 and 3 circuits respectively.}
\end{figure*}

When considering the combined effects of decoherence, stochastic fluctuations due to finite sampling and limited number of trial states, the advantages of using more entanglers may not be apparent anymore. This is the case for many of the molecular Hamiltonians discussed in this paper, whose energies are well approximated by separable states prepared using low-depth circuits. In Fig.~\ref{ExpDepth} we show the experimental optimization for different depths, $d=0,1,2$, for the Hamiltonian of LiH at the bond distance, compared with 100 outcomes of numerical simulations. The numerical histograms in Fig.~\ref{ExpDepth}b show large overlap between final energy distributions for $d=0,1,2$, confirmed by the experiments presented in Fig.~\ref{ExpDepth}a. 
This overlap between outcomes of optimizations with different entanglers appear for most of the molecular Hamiltonians. In contrast, for the interacting spin Hamiltonians discussed in Fig. 4 of the main text, significantly better estimates are obtained with $d=1,2,3$ circuits than $d=0$ circuits.

\cleardoublepage

\begin{longtable}{|c|c|c|c|c|c|c|c|c|}
\caption{\label{HamTable}The H$_2$, LiH and BeH$_2$ Hamiltonians at the bond distance. Listed are all the Pauli operators, grouped in the different TPB sets, with the corresponding coefficients, not taking into account for the energy shifts due to the filling of inner orbitals and the Coulomb repulsion between nuclei. X,Y,Z,I here stand for the Pauli matrices $\sigma^x$, $\sigma^y$, $\sigma^z$ and the identity operator on a single qubit subspace, respectively. There are 2,25,44 TPB sets for H$_2$, LiH and BeH$_2$, respectively with 4, 99 and 164 Pauli terms in total.  }
\\
\multicolumn{9}{c}{} \\[2ex]\\
\multicolumn{9}{c} {\large H$_2$ at bond distance}\\[2ex]\\
\hline
\multicolumn{4}{|c|}{
\parbox{1.9cm}{\centering\vspace{.5\baselineskip}
ZZ\\
0.011280\\
ZI\\
0.397936\\ 
IZ\\
0.397936\\ 
\vspace{.5\baselineskip}} 
}
&
\multicolumn{5}{|c|}{
\parbox{1.9cm}{\centering\vspace{.5\baselineskip}
 XX\\
0.180931
\vspace{.5\baselineskip}} 
}\\
\hline\\
\multicolumn{9}{c}{} \\[2ex]\\
\multicolumn{9}{c} {\large LiH at bond distance}\\[2ex]\\  
\hline
\parbox{1.9cm}{\centering\vspace{.5\baselineskip}
ZIII\\
-0.096022\\ 
ZZII\\
-0.206128\\ 
IZII\\
0.364746\\ 
IIZI\\
0.096022\\ 
IIZZ\\
-0.206128\\ 
IIIZ\\
-0.364746\\ 
ZIZI\\
-0.145438\\ 
ZIZZ\\
0.056040\\ 
ZIIZ\\
0.110811\\ 
ZZZI\\
-0.056040\\ 
ZZZZ\\
0.080334\\ 
ZZIZ\\
0.063673\\ 
IZZI\\
0.110811\\ 
IZZZ\\
-0.063673\\ 
IZIZ\\
-0.095216\\ 
\vspace{.5\baselineskip}}
&
\parbox{1.9cm}{\centering\vspace{.5\baselineskip}
XZII\\
-0.012585\\ 
XIII\\
0.012585\\ 
IIXZ\\
0.012585\\ 
IIXI\\
0.012585\\ 
XZXZ\\
-0.002667\\ 
XZXI\\
-0.002667\\ 
XIXZ\\
0.002667\\ 
XIXI\\
0.002667\\ 
XZIZ\\
0.007265\\ 
XIIZ\\
-0.007265\\ 
IZXZ\\
0.007265\\ 
IZXI\\
0.007265\\ 
\vspace{.5\baselineskip}}
&
\parbox{1.9cm}{\centering\vspace{.5\baselineskip}
XXII\\
-0.029640\\ 
IXII\\
0.002792\\ 
IIXX\\
-0.029640\\ 
IIIX\\
0.002792\\ 
XIXX\\
-0.008195\\ 
XIIX\\
-0.001271\\ 
XXXI\\
-0.008195\\ 
XXXX\\
0.028926\\ 
XXIX\\
0.007499\\ 
IXXI\\
-0.001271\\ 
IXXX\\
0.007499\\ 
IXIX\\
0.009327\\ 
\vspace{.5\baselineskip}}
&
\parbox{1.9cm}{\centering\vspace{.5\baselineskip}
YYII\\
0.029640\\ 
IIYY\\
0.029640\\ 
YYYY\\
0.028926\\ 
\vspace{.5\baselineskip}}
&
\parbox{1.9cm}{\centering\vspace{.5\baselineskip}
ZXII\\
0.002792\\ 
IIZX\\
-0.002792\\ 
ZIZX\\
-0.016781\\ 
ZIIX\\
0.016781\\ 
ZXZI\\
-0.016781\\ 
IXZI\\
-0.016781\\ 
ZXZX\\
-0.009327\\ 
ZXIX\\
0.009327\\ 
IXZX\\
-0.009327\\ 
\vspace{.5\baselineskip}}
&
\parbox{1.9cm}{\centering\vspace{.5\baselineskip}
ZIXZ\\
-0.011962\\ 
ZIXI\\
-0.011962\\ 
ZZXZ\\
0.000247\\ 
ZZXI\\
0.000247\\ 
\vspace{.5\baselineskip}}
&
\parbox{1.9cm}{\centering\vspace{.5\baselineskip}
ZIXX\\
0.039155\\ 
ZZXX\\
-0.002895\\ 
ZZIX\\
-0.009769\\ 
IZXX\\
-0.024280\\ 
IZIX\\
-0.008025\\ 
\vspace{.5\baselineskip}}
&
\parbox{1.9cm}{\centering\vspace{.5\baselineskip}
ZIYY\\
-0.039155\\ 
ZZYY\\
0.002895\\ 
IZYY\\
0.024280\\ 
\vspace{.5\baselineskip}}
&
\parbox{1.9cm}{\centering\vspace{.5\baselineskip}
XZZI\\
-0.011962\\ 
XIZI\\
0.011962\\ 
XZZZ\\
-0.000247\\ 
XIZZ\\
0.000247\\ 
\vspace{.5\baselineskip}}\\
\hline
\parbox{1.9cm}{\centering\vspace{.5\baselineskip}
XZXX\\
0.008195\\ 
XZIX\\
0.001271\\ 
\vspace{.5\baselineskip}}
&
\parbox{1.9cm}{\centering\vspace{.5\baselineskip}
XZYY\\
-0.008195\\ 
XIYY\\
0.008195\\ 
\vspace{.5\baselineskip}}
&
\parbox{1.9cm}{\centering\vspace{.5\baselineskip}
XZZX\\
-0.001271\\ 
XIZX\\
0.001271\\ 
IZZX\\
0.008025\\ 
\vspace{.5\baselineskip}}
&
\parbox{1.9cm}{\centering\vspace{.5\baselineskip}
XXZI\\
-0.039155\\ 
XXZZ\\
-0.002895\\ 
XXIZ\\
0.024280\\ 
IXZZ\\
-0.009769\\ 
IXIZ\\
0.008025\\ 
\vspace{.5\baselineskip}}
&
\parbox{1.9cm}{\centering\vspace{.5\baselineskip}
YYZI\\
0.039155\\ 
YYZZ\\
0.002895\\ 
YYIZ\\
-0.024280\\ 
\vspace{.5\baselineskip}}
&
\parbox{1.9cm}{\centering\vspace{.5\baselineskip}
XXXZ\\
-0.008195\\ 
IXXZ\\
-0.001271\\ 
\vspace{.5\baselineskip}}
&
\parbox{1.9cm}{\centering\vspace{.5\baselineskip}
YYXZ\\
0.008195\\ 
YYXI\\
0.008195\\ 
\vspace{.5\baselineskip}}
&
\parbox{1.9cm}{\centering\vspace{.5\baselineskip}
XXYY\\
-0.028926\\ 
IXYY\\
-0.007499\\ 
\vspace{.5\baselineskip}}
&
\parbox{1.9cm}{\centering\vspace{.5\baselineskip}
YYXX\\
-0.028926\\ 
YYIX\\
-0.007499\\ 
\vspace{.5\baselineskip}}\\
\hline
\parbox{1.9cm}{\centering\vspace{.5\baselineskip}
XXZX\\
-0.007499\\ 
\vspace{.5\baselineskip}}
&
\parbox{1.9cm}{\centering\vspace{.5\baselineskip}
YYZX\\
0.007499\\ 
\vspace{.5\baselineskip}}
&
\parbox{1.9cm}{\centering\vspace{.5\baselineskip}
ZZZX\\
0.009769\\ 
\vspace{.5\baselineskip}}
&
\parbox{1.9cm}{\centering\vspace{.5\baselineskip}
ZXXZ\\
-0.001271\\ 
ZXXI\\
-0.001271\\ 
ZXIZ\\
0.008025\\ 
\vspace{.5\baselineskip}}
&
\parbox{1.9cm}{\centering\vspace{.5\baselineskip}
ZXXX\\
0.007499\\ 
\vspace{.5\baselineskip}}
&
\parbox{1.9cm}{\centering\vspace{.5\baselineskip}
ZXYY\\
-0.007499\\ 
\vspace{.5\baselineskip}}
&
\parbox{1.9cm}{\centering\vspace{.5\baselineskip}
ZXZZ\\
-0.009769\\ 
\vspace{.5\baselineskip}}
&
&\\
\hline
\multicolumn{9}{c}{} \\[2ex]\\
\multicolumn{9}{c} {\large BeH$_2$ at bond distance}\\[2ex]\\			
\hline
\multicolumn{1}{|c}{
\parbox{1.9cm}{\centering\vspace{.5\baselineskip}
ZIIIII\\
-0.143021\\ 
ZZIIII\\
0.104962\\ 
IZZIII\\
0.038195\\ 
IIZIII\\
-0.325651\\ 
IIIZII\\
-0.143021\\ 
IIIZZI\\
0.104962\\ 
IIIIZZ\\
0.038195\\ 
IIIIIZ\\
-0.325651\\ 
IZIIII\\
0.172191\\ 
ZZZIII\\
0.174763\\ 
ZIZIII\\
0.136055\\ 
ZIIZII\\
0.116134\\ 
ZIIZZI\\
0.094064\\ 
ZIIIZZ\\
0.099152\\ 
ZIIIIZ\\
0.123367\\
\vspace{.5\baselineskip}}
}
&
\multicolumn{1}{c|}{
\parbox{1.9cm}{\centering\vspace{.5\baselineskip}
ZZIZII\\
0.094064\\ 
ZZIZZI\\
0.098003\\ 
ZZIIZZ\\
0.102525\\ 
ZZIIIZ\\
0.097795\\ 
IZZZII\\
0.099152\\ 
IZZZZI\\
0.102525\\ 
IZZIZZ\\
0.112045\\ 
IZZIIZ\\
0.105708\\ 
IIZZII\\
0.123367\\ 
IIZZZI\\
0.097795\\ 
IIZIZZ\\
0.105708\\ 
IIZIIZ\\
0.133557\\ 
IIIIZI\\
0.172191\\ 
IIIZZZ\\
0.174763\\ 
IIIZIZ\\
0.136055\\
\vspace{.5\baselineskip}}
}
&
\multicolumn{1}{|c}{
\parbox{1.9cm}{\centering\vspace{.5\baselineskip}
XZIIII\\
0.059110\\ 
XIIIII\\
-0.059110\\ 
IZXIII\\
0.161019\\ 
IIXIII\\
-0.161019\\ 
IIIXZI\\
0.059110\\ 
IIIXII\\
-0.059110\\ 
IIIIZX\\
0.161019\\ 
IIIIIX\\
-0.161019\\ 
XIXIII\\
-0.038098\\ 
XZXIII\\
-0.003300\\ 
XZIXZI\\
0.013745\\ 
XZIXII\\
-0.013745\\ 
XIIXZI\\
-0.013745\\ 
XIIXII\\
0.013745\\ 
\vspace{.5\baselineskip}}
}
&
\multicolumn{1}{c|}{
\parbox{1.9cm}{\centering\vspace{.5\baselineskip}
XZIIZX\\
0.011986\\ 
XZIIIX\\
-0.011986\\ 
XIIIZX\\
-0.011986\\ 
XIIIIX\\
0.011986\\ 
IZXXZI\\
0.011986\\ 
IZXXII\\
-0.011986\\ 
IIXXZI\\
-0.011986\\ 
IIXXII\\
0.011986\\ 
IZXIZX\\
0.013836\\ 
IZXIIX\\
-0.013836\\ 
IIXIZX\\
-0.013836\\ 
IIXIIX\\
0.013836\\ 
IIIXIX\\
-0.038098\\ 
IIIXZX\\
-0.003300\\ 
\vspace{.5\baselineskip}}
}
&
\parbox{1.9cm}{\centering\vspace{.5\baselineskip}
ZZXIII\\
-0.002246\\ 
ZIXIII\\
0.002246\\ 
ZIIXZI\\
0.014815\\ 
ZIIXII\\
-0.014815\\ 
ZIIIZX\\
0.009922\\ 
ZIIIIX\\
-0.009922\\ 
ZZIXZI\\
-0.002038\\ 
ZZIXII\\
0.002038\\ 
ZZIIZX\\
-0.007016\\ 
ZZIIIX\\
0.007016\\ 
\vspace{.5\baselineskip}}
&
\parbox{1.9cm}{\centering\vspace{.5\baselineskip}
XIZIII\\
-0.006154\\ 
XZZIII\\
0.006154\\ 
XZIZII\\
0.014815\\ 
XIIZII\\
-0.014815\\ 
XZIZZI\\
-0.002038\\ 
XIIZZI\\
0.002038\\ 
XZIIZZ\\
0.001124\\ 
XIIIZZ\\
-0.001124\\ 
XZIIIZ\\
0.017678\\ 
XIIIIZ\\
-0.017678\\ 
\vspace{.5\baselineskip}}
&
\parbox{1.9cm}{\centering\vspace{.5\baselineskip}
YIYIII\\
-0.041398\\ 
YYIXXZ\\
0.011583\\ 
YYIIXI\\
-0.011094\\ 
IYYXXZ\\
0.010336\\ 
IYYIXI\\
-0.005725\\ 
IIIXIZ\\
-0.006154\\ 
\vspace{.5\baselineskip}}
&
\parbox{1.9cm}{\centering\vspace{.5\baselineskip}
XXZXXZ\\
0.011583\\ 
XXZIXI\\
-0.011094\\ 
IXIXXZ\\
-0.011094\\ 
IXIIXI\\
0.026631\\ 
IIZXII\\
-0.017678\\ 
\vspace{.5\baselineskip}}
&
\parbox{1.9cm}{\centering\vspace{.5\baselineskip}
XXZYYI\\
0.011583\\ 
XXZIYY\\
0.010336\\ 
IXIYYI\\
-0.011094\\ 
IXIIYY\\
-0.005725\\ 
IIIYIY\\
-0.041398\\ 
\vspace{.5\baselineskip}}\\									
\hline
\parbox{1.9cm}{\centering\vspace{.5\baselineskip}
YYIYYI\\
0.011583\\ 
YYIIYY\\
0.010336\\ 
IYYYYI\\
0.010336\\ 
IYYIYY\\
0.010600\\ 
\vspace{.5\baselineskip}}
&
\parbox{1.9cm}{\centering\vspace{.5\baselineskip}
XXZXXX\\
0.024909\\ 
IXIXXX\\
-0.031035\\ 
IIZIIX\\
-0.010064\\ 
\vspace{.5\baselineskip}}
&
\parbox{1.9cm}{\centering\vspace{.5\baselineskip}
XXZYXY\\
0.024909\\ 
IXIYXY\\
-0.031035\\ 
\vspace{.5\baselineskip}}
&
\parbox{1.9cm}{\centering\vspace{.5\baselineskip}
YYIXXX\\
0.024909\\ 
IYYXXX\\
0.021494\\ 
\vspace{.5\baselineskip}}
&
\parbox{1.9cm}{\centering\vspace{.5\baselineskip}
YYIYXY\\
0.024909\\ 
IYYYXY\\
0.021494\\ 
\vspace{.5\baselineskip}}
&
\parbox{1.9cm}{\centering\vspace{.5\baselineskip}
XXZZXZ\\
0.011094\\ 
IXIZXZ\\
-0.026631\\ 
\vspace{.5\baselineskip}}
&
\parbox{1.9cm}{\centering\vspace{.5\baselineskip}
YYIZXZ\\
0.011094\\ 
IYYZXZ\\
0.005725\\ 
\vspace{.5\baselineskip}}
&
\parbox{1.9cm}{\centering\vspace{.5\baselineskip}
XXZZXX\\
0.010336\\ 
IXIZXX\\
-0.005725\\ 
IIIZIX\\
0.002246\\ 
\vspace{.5\baselineskip}}
&
\parbox{1.9cm}{\centering\vspace{.5\baselineskip}
YYIZXX\\
0.010336\\ 
IYYZXX\\
0.010600\\ 
\vspace{.5\baselineskip}}\\									
\hline
\parbox{1.9cm}{\centering\vspace{.5\baselineskip}
XXXXXZ\\
0.024909\\ 
XXXIXI\\
-0.031035\\ 
IIXIIZ\\
-0.010064\\ 
\vspace{.5\baselineskip}}
&
\parbox{1.9cm}{\centering\vspace{.5\baselineskip}
XXXYYI\\
0.024909\\ 
XXXIYY\\
0.021494\\ 
\vspace{.5\baselineskip}}
&
\parbox{1.9cm}{\centering\vspace{.5\baselineskip}
YXYXXZ\\
0.024909\\ 
YXYIXI\\
-0.031035\\ 
\vspace{.5\baselineskip}}
&
\parbox{1.9cm}{\centering\vspace{.5\baselineskip}
YXYYYI\\
0.024909\\ 
YXYIYY\\
0.021494\\ 
\vspace{.5\baselineskip}}
&
\parbox{1.9cm}{\centering\vspace{.5\baselineskip}
XXXXXX\\
0.063207\\ 
\vspace{.5\baselineskip}}
&
\parbox{1.9cm}{\centering\vspace{.5\baselineskip}
XXXYXY\\
0.063207\\ 
\vspace{.5\baselineskip}}
&
\parbox{1.9cm}{\centering\vspace{.5\baselineskip}
YXYXXX\\
0.063207\\ 
\vspace{.5\baselineskip}}
&
\parbox{1.9cm}{\centering\vspace{.5\baselineskip}
YXYYXY\\
0.063207\\ 
\vspace{.5\baselineskip}}
&
\parbox{1.9cm}{\centering\vspace{.5\baselineskip}
XXXZXZ\\
0.031035\\ 
IIXZII\\
-0.009922\\ 
\vspace{.5\baselineskip}}\\		
\hline													
\parbox{1.9cm}{\centering\vspace{.5\baselineskip}
YXYZXZ\\
0.031035\\ 
\vspace{.5\baselineskip}}
&
\parbox{1.9cm}{\centering\vspace{.5\baselineskip}
XXXZXX\\
0.021494\\ 
\vspace{.5\baselineskip}}
&
\parbox{1.9cm}{\centering\vspace{.5\baselineskip}
YXYZXX\\
0.021494\\ 
\vspace{.5\baselineskip}}
&
\parbox{1.9cm}{\centering\vspace{.5\baselineskip}
ZXZXXZ\\
0.011094\\ 
ZXZIXI\\
-0.026631\\ 
\vspace{.5\baselineskip}}
&
\parbox{1.9cm}{\centering\vspace{.5\baselineskip}
ZXZYYI\\
0.011094\\ 
ZXZIYY\\
0.005725\\ 
\vspace{.5\baselineskip}}
&
\parbox{1.9cm}{\centering\vspace{.5\baselineskip}
ZXZXXX\\
0.031035\\ 
\vspace{.5\baselineskip}}
&
\parbox{1.9cm}{\centering\vspace{.5\baselineskip}
ZXZYXY\\
0.031035\\ 
\vspace{.5\baselineskip}}
&
\parbox{1.9cm}{\centering\vspace{.5\baselineskip}
ZXZZXZ\\
0.026631\\ 
\vspace{.5\baselineskip}}
&
\parbox{1.9cm}{\centering\vspace{.5\baselineskip}
ZXZZXX\\
0.005725\\ 
\vspace{.5\baselineskip}}\\		
\hline						
\parbox{1.9cm}{\centering\vspace{.5\baselineskip}
ZXXXXZ\\
0.010336\\ 
ZXXIXI\\
-0.005725\\ 
\vspace{.5\baselineskip}}
&
\parbox{1.9cm}{\centering\vspace{.5\baselineskip}
ZXXYYI\\
0.010336\\ 
ZXXIYY\\
0.010600\\ 
\vspace{.5\baselineskip}}
&
\parbox{1.9cm}{\centering\vspace{.5\baselineskip}
ZXXXXX\\
0.021494\\ 
\vspace{.5\baselineskip}}
&
\parbox{1.9cm}{\centering\vspace{.5\baselineskip}
ZXXYXY\\
0.021494\\ 
\vspace{.5\baselineskip}}
&
\parbox{1.9cm}{\centering\vspace{.5\baselineskip}
ZXXZXZ\\
0.005725\\ 
\vspace{.5\baselineskip}}
&
\parbox{1.9cm}{\centering\vspace{.5\baselineskip}
ZXXZXX\\
0.010600\\ 
\vspace{.5\baselineskip}}
&
\parbox{1.9cm}{\centering\vspace{.5\baselineskip}
IZZXZI\\
0.001124\\ 
IZZXII\\
-0.001124\\ 
IZZIZX\\
-0.007952\\ 
IZZIIX\\
0.007952\\ 
IIZXZI\\
0.017678\\ 
IIZIZX\\
0.010064\\ 
\vspace{.5\baselineskip}}
&
\parbox{1.9cm}{\centering\vspace{.5\baselineskip}
IZXZII\\
0.009922\\ 
IZXZZI\\
-0.007016\\ 
IIXZZI\\
0.007016\\ 
IZXIZZ\\
-0.007952\\ 
IIXIZZ\\
0.007952\\ 
IZXIIZ\\
0.010064\\ 
\vspace{.5\baselineskip}}
&
\parbox{1.9cm}{\centering\vspace{.5\baselineskip}
IIIZZX\\
-0.002246\\ 
\vspace{.5\baselineskip}}\\								
\hline
\parbox{1.9cm}{\centering\vspace{.5\baselineskip}
IIIXZZ \\
0.006154 \\
\vspace{.5\baselineskip}}
&
&
&
&
&
&
&
&\\
\hline
\end{longtable}

\end{document}